%% file: main.tex
\documentclass[final,5p,times,twocolumn]{elsarticle}
\usepackage{amssymb}
\usepackage{amsmath,amsfonts}
\usepackage{algorithmic}
\usepackage{algorithm}
\usepackage{array}
\usepackage[caption=false,font=normalsize,labelfont=sf,textfont=sf]{subfig}
\usepackage{textcomp}
\usepackage[hyphens]{url}
\makeatletter
\g@addto@macro\UrlBreaks{%
  \do\a\do\b\do\c\do\d\do\e\do\f\do\g\do\h\do\i\do\j\do\k\do\l\do\m%
  \do\n\do\o\do\p\do\q\do\r\do\s\do\t\do\u\do\v\do\w\do\x\do\y\do\z%
  \do\A\do\B\do\C\do\D\do\E\do\F\do\G\do\H\do\I\do\J\do\K\do\L\do\M%
  \do\N\do\O\do\P\do\Q\do\R\do\S\do\T\do\U\do\V\do\W\do\X\do\Y\do\Z%
  \do\0\do\1\do\2\do\3\do\4\do\5\do\6\do\7\do\8\do\9%
  \do\-\do\_\do\.\do\=\do\&\do\?\do\#\do\%\do\+\do\~\do\@\do\:\do\,}
\makeatother
\usepackage{verbatim}
\usepackage{graphicx}
\usepackage{multirow}
\usepackage{booktabs}
\usepackage{xspace}
\usepackage[dvipsnames]{xcolor}
\usepackage[many]{tcolorbox}
\definecolor{customcite}{HTML}{b83b5e}
\definecolor{customlink}{HTML}{07689f}
\definecolor{customurl}{HTML}{11999e}
\usepackage{hyperref}

\hypersetup{
  colorlinks = true,
  breaklinks = true,
  linkcolor = customlink,
  anchorcolor = purple,
  citecolor = customcite,
  filecolor = purple,
  urlcolor = customurl
}
\newtcolorbox{findingbox}{
    colback=gray!10,
    colframe=gray!50,
    boxrule=0.5pt,
    arc=1mm,
    boxsep=1.5mm,
    left=1.7mm,right=1.7mm,top=1mm,bottom=1mm,
    fontupper=\itshape
}

\newtcolorbox{notebox}[1][]{
colback=ForestGreen!15, 
colframe=ForestGreen!60,
    boxrule=0.5pt,
    arc=1.5mm,
    boxsep=1.2mm,
    left=2mm,right=2mm,top=2mm,bottom=2mm,
    #1
}
\newtcolorbox{answerbox}{
  enhanced,
  left=1.7mm,
  right=1.7mm,
  top=1.7mm,
  bottom=1.7mm,
  colback=gray!10,  
  colframe=gray!90, 
  boxrule=0pt,      
  leftrule=3pt,     
  sharp corners,    
  breakable         
}

\newcounter{insightcounter}
\newcommand{\observation}[1]{%
    \stepcounter{insightcounter}%
    \begin{tcolorbox}[colframe=black, boxrule=0.8pt, arc=1mm]
    \textbf{Observation \theinsightcounter:} #1
    \end{tcolorbox}
}

\newcommand{\paratitle}[1]{\noindent\textbf{#1.}\quad}
\newcommand{\etal}{{\textit{et al.}}\xspace}

\input{numbers}

\journal{Blockchain: Research and Applications}

\begin{document}

\begin{frontmatter}

\title{Web3 \texorpdfstring{$\times$}{×} AI Agents: Landscape, Integrations, and Foundational Challenges}

\author[sysu]{Yiming Shen}
\ead{shenym7@mail2.sysu.edu.cn}

\author[zjucs]{Jiashuo Zhang}

\author[sysu]{Zhenzhe Shao}

\author[uestc]{Wenxuan Luo}

\author[sysu]{Yanlin Wang}

\author[uestc]{Ting Chen}

\author[zju,zjubj]{Lingfeng Bao}

\author[zju,zjubj]{Jiachi Chen\corref{cor1}}
\ead{chenjiachi@zju.edu.cn}

\cortext[cor1]{Corresponding author.}

\affiliation[sysu]{organization={School of Software Engineering, Sun Yat-sen University},
            city={Zhuhai},
            state={Guangdong},
            country={China}}

\affiliation[zjucs]{organization={College of Computer Science and Technology, Zhejiang University},
            city={Hangzhou},
            state={Zhejiang},
            country={China}}

\affiliation[uestc]{organization={School of Computer Science and Engineering (School of Cyber Security), University of Electronic Science and Technology of China},
            city={Chengdu},
            state={Sichuan},
            country={China}}

\affiliation[zju]{organization={The State Key Laboratory of Blockchain and Data Security, Zhejiang University},
            city={Hangzhou},
            state={Zhejiang},
            country={China}}

\affiliation[zjubj]{organization={Hangzhou High-Tech Zone (Binjiang) Institute of Blockchain and Data Security},
            city={Hangzhou},
            state={Zhejiang},
            country={China}}

\begin{abstract}

The convergence of Web3 technologies and AI agents is a rapidly evolving frontier that is reshaping decentralized ecosystems. This paper presents \textbf{\textit{the first comprehensive landscape analysis}} of the intersection between Web3 and AI agents, examining five critical dimensions: \textit{landscape}, \textit{economics}, \textit{governance}, \textit{security}, and \textit{trust mechanisms}. Through an analysis of \nPopCount capitalized Web3-AI agent tokens at a \nDateTwo cutoff, compared with a cohort of \nCohCount tokens collected at \nDateOne, we develop a taxonomy and systematically map the current market landscape (RQ1), identifying distinct patterns in project distribution and capitalization. We complement the token frame with on-chain measurement of the agent standards layer: \nErcRegistrations ERC-8004 identity registrations across \nErcChains chains and \nXFourSettlements x402 settlements on \nXFourChain. We then investigate four integration domains: AI agents in decentralized finance (RQ2), governance (RQ3), security (RQ4), and Web3 trust mechanisms for agent reliability (RQ5). Infrastructure dominates the landscape in both project count and capitalization, value is highly concentrated (Gini \nPopGini), and the archived cohort lost \nCohChange in capitalization over \nWindow despite token population growth. The agent standards layer shows adoption at scale, but economic throughput remains limited. Across integration domains, AI agents extend DeFi trading and make continuous security auditing affordable, while agents in governance remain largely a blueprint. These findings outline critical considerations for future research toward robust, intelligent, and trustworthy decentralized systems with AI agent interactions.

\end{abstract}

\begin{keyword}
Web3 \sep Blockchain \sep Artificial Intelligence (AI) \sep LLM \sep Agent
\end{keyword}

\end{frontmatter}

\section{Introduction}
\input{tex/01-introduction}

\section{Background}\label{sec:background}
\input{tex/02-background}

\section{Methodology}

\input{tex/03-methodology}

\section{Answer to RQ1: Landscape of Web3 AI Agent}\label{sec:rq1}
\input{tex/04-rq1}

\section{Answer to RQ2: AI Agents in Decentralized Finance}
\input{tex/05-rq2}

\section{Answer to RQ3: AI Agents Enhancing Web3 Governance}
\input{tex/06-rq3}

\section{Answer to RQ4: AI Agents for Web3 Security}
\input{tex/07-rq4}

\section{Answer to RQ5: Web3 Trust Mechanisms for AI Agents}
\input{tex/08-rq5}

\section{Discussion}
\input{tex/09-discussion}

\section{Related Work}
\input{tex/10-related_work}

\section{Conclusion}
\input{tex/11-conclusion}

\section*{Declaration of Competing Interest}

The authors declare that they have no known competing financial interests or
personal relationships that could have appeared to influence the work reported
in this paper.

\bibliographystyle{elsarticle-num}
\bibliography{reference}

\end{document}

%% file: numbers.tex
\newcommand{\nDateOne}      {6 April 2025\xspace}
\newcommand{\nDateTwo}      {1 September 2026\xspace}
\newcommand{\nWindow}       {seventeen months\xspace}
\newcommand{\nAliveFloor}   {\$1 million\xspace}
\newcommand{\nStaleDays}    {180\xspace}

\newcommand{\nPopCount}                 {979\xspace}
\newcommand{\nPopMcap}                  {\$16.03 billion\xspace}

\newcommand{\nPopMedian}                {\$234{,}397\xspace}
\newcommand{\nPopGini}                  {0.938\xspace}
\newcommand{\nPopTopPctCount}           {10\xspace}
\newcommand{\nPopTopPctShare}           {57.8\%\xspace}

\newcommand{\nPopTopTwentiethShare}     {82.7\%\xspace}

\newcommand{\nPopTopDecileShare}        {91.1\%\xspace}
\newcommand{\nPopBottomShare}           {8.9\%\xspace}

\newcommand{\nPopSubHundredKShare}      {39.2\%\xspace}

\newcommand{\nPopHundredKToMShare}      {27.4\%\xspace}

\newcommand{\nPopAboveHundredMCount}    {25\xspace}
\newcommand{\nPopAboveHundredMShare}    {2.6\%\xspace}

\newcommand{\nScrFrame}                {1{,}837\xspace}

\newcommand{\nScrNotCapitalised}       {545\xspace}
\newcommand{\nScrDerivative}           {3\xspace}
\newcommand{\nScrDerivativeValue}      {\$1.64 billion\xspace}
\newcommand{\nScrDuplicate}            {12\xspace}
\newcommand{\nScrDuplicateGroups}      {13\xspace}
\newcommand{\nScrMeme}                 {216\xspace}
\newcommand{\nScrMemeValue}            {\$497 million\xspace}
\newcommand{\nScrMechanical}           {1{,}061\xspace}
\newcommand{\nScrMechanicalMcap}       {\$16.26 billion\xspace}
\newcommand{\nScrNoDescription}        {82\xspace}
\newcommand{\nScrNoDescriptionValue}   {\$230 million\xspace}
\newcommand{\nScrNoDescriptionShare}   {1.4\%\xspace}

\newcommand{\nClsCount}            {855\xspace}
\newcommand{\nClsMcap}             {\$15.46 billion\xspace}

\newcommand{\nClsOutOfScope}       {87\xspace}
\newcommand{\nClsUnclassifiable}   {37\xspace}
\newcommand{\nClsRule}             {318\xspace}
\newcommand{\nClsRead}             {537\xspace}

\newcommand{\nClsMulti}            {120\xspace}
\newcommand{\nClsMultiShare}       {14.0\%\xspace}
\newcommand{\nClsMultiMcap}        {\$1.49 billion\xspace}
\newcommand{\nClsMultiMcapShare}   {9.6\%\xspace}

\newcommand{\nClsIncubCount}      {213\xspace}
\newcommand{\nClsIncubMcap}       {\$2.44 billion\xspace}
\newcommand{\nClsIncubMean}       {\$11.5 million\xspace}
\newcommand{\nClsIncubShare}      {15.8\%\xspace}
\newcommand{\nClsInfraCount}      {305\xspace}
\newcommand{\nClsInfraMcap}       {\$11.74 billion\xspace}
\newcommand{\nClsInfraMean}       {\$38.5 million\xspace}
\newcommand{\nClsInfraShare}      {75.9\%\xspace}
\newcommand{\nClsFinCount}        {278\xspace}
\newcommand{\nClsFinMcap}         {\$1.28 billion\xspace}
\newcommand{\nClsFinMean}         {\$4.6 million\xspace}
\newcommand{\nClsFinShare}        {8.3\%\xspace}
\newcommand{\nClsCreativeCount}   {137\xspace}
\newcommand{\nClsCreativeMcap}    {\$591 million\xspace}
\newcommand{\nClsCreativeMean}    {\$4.3 million\xspace}
\newcommand{\nClsCreativeShare}   {3.8\%\xspace}
\newcommand{\nClsFinCountShare}   {29.8\%\xspace}

\newcommand{\nClsBuilderCount}     {160\xspace}
\newcommand{\nClsBuilderMcap}      {\$1.90 billion\xspace}

\newcommand{\nClsLaunchpadCount}   {69\xspace}

\newcommand{\nClsLaunchpadMean}    {\$17.2 million\xspace}

\newcommand{\nClsProtocolCount}    {226\xspace}

\newcommand{\nClsProtocolMean}     {\$50.4 million\xspace}

\newcommand{\nClsTrustCount}       {70\xspace}

\newcommand{\nClsCopilotCount}     {27\xspace}
\newcommand{\nClsCopilotMcap}      {\$30.8 million\xspace}
\newcommand{\nClsCopilotMean}      {\$1.1 million\xspace}

\newcommand{\nClsDeFAICount}       {185\xspace}

\newcommand{\nClsDeFAIMean}        {\$5.1 million\xspace}

\newcommand{\nClsAnalyticsCount}   {107\xspace}

\newcommand{\nClsGameCount}        {64\xspace}

\newcommand{\nClsContentCount}     {68\xspace}

\newcommand{\nClsRWACount}         {8\xspace}

\newcommand{\nClsFacetChainCount}     {29\xspace}
\newcommand{\nClsFacetChainMcap}      {\$4.64 billion\xspace}
\newcommand{\nClsFacetChainMean}      {\$160 million\xspace}
\newcommand{\nClsFacetChainShare}     {40.7\%\xspace}
\newcommand{\nClsFacetComputeCount}   {89\xspace}

\newcommand{\nClsFacetComputeMean}    {\$31.3 million\xspace}

\newcommand{\nClsFacetModelCount}     {41\xspace}

\newcommand{\nClsFacetModelMean}      {\$67.5 million\xspace}

\newcommand{\nClsFacetDataCount}      {67\xspace}

\newcommand{\nClsFacetDataMean}       {\$17.8 million\xspace}

\newcommand{\nClsChains}                     {56\xspace}
\newcommand{\nClsDeployments}                {1{,}180\xspace}
\newcommand{\nClsListed}                     {840\xspace}
\newcommand{\nClsListedShare}                {98.2\%\xspace}
\newcommand{\nClsMultichain}                 {187\xspace}
\newcommand{\nClsMultichainShare}            {22.3\%\xspace}
\newcommand{\nClsChainBaseCount}             {326\xspace}
\newcommand{\nClsChainBaseMcap}              {\$6.76 billion\xspace}
\newcommand{\nClsChainBaseMean}              {\$20.7 million\xspace}

\newcommand{\nClsChainBaseCountShare}        {27.6\%\xspace}
\newcommand{\nClsChainEthereumCount}         {198\xspace}
\newcommand{\nClsChainEthereumMcap}          {\$5.87 billion\xspace}
\newcommand{\nClsChainEthereumMean}          {\$29.6 million\xspace}

\newcommand{\nClsChainEthereumCountShare}    {16.8\%\xspace}
\newcommand{\nClsChainSolanaCount}           {249\xspace}
\newcommand{\nClsChainSolanaMcap}            {\$2.21 billion\xspace}
\newcommand{\nClsChainSolanaMean}            {\$8.9 million\xspace}

\newcommand{\nClsChainSolanaCountShare}      {21.1\%\xspace}
\newcommand{\nClsChainBNBCount}              {136\xspace}
\newcommand{\nClsChainBNBMcap}               {\$2.70 billion\xspace}
\newcommand{\nClsChainBNBMean}               {\$19.9 million\xspace}

\newcommand{\nClsChainBNBCountShare}         {11.5\%\xspace}
\newcommand{\nClsChainBittensorCount}        {63\xspace}

\newcommand{\nClsChainRobinhoodCount}        {34\xspace}

\newcommand{\nClsChainRobinhoodMean}         {\$80.5 million\xspace}

\newcommand{\nClsChainArbitrumCount}         {33\xspace}

\newcommand{\nClsChainPolygonCount}          {21\xspace}

\newcommand{\nClsPairOne}         {18\xspace}
\newcommand{\nClsPairOneName}     {\textit{Protocol} with \textit{Trust}\xspace}
\newcommand{\nClsPairTwo}         {16\xspace}
\newcommand{\nClsPairTwoName}     {\textit{Builder} with \textit{Launchpad}\xspace}
\newcommand{\nClsPairThree}       {15\xspace}
\newcommand{\nClsPairThreeName}   {\textit{Builder} with \textit{DeFAI}\xspace}
\newcommand{\nClsPairFour}        {14\xspace}
\newcommand{\nClsPairFourName}    {\textit{Analytics} with \textit{DeFAI}\xspace}
\newcommand{\nActCount}           {331\xspace}

\newcommand{\nActIncubRate}       {60\%\xspace}

\newcommand{\nActInfraRate}       {71\%\xspace}

\newcommand{\nActFinRate}         {56\%\xspace}

\newcommand{\nActCreativeRate}    {46\%\xspace}

\newcommand{\nTrendSpan}      {eighteen months\xspace}
\newcommand{\nTrendFirst}     {2025-03\xspace}
\newcommand{\nTrendLast}      {2026-08\xspace}
\newcommand{\nTrendLiveOne}   {387\xspace}
\newcommand{\nTrendLiveTwo}   {854\xspace}
\newcommand{\nTrendMcapOne}   {\$23.12 billion\xspace}
\newcommand{\nTrendMcapTwo}   {\$15.19 billion\xspace}

\newcommand{\nCohCount}            {275\xspace}
\newcommand{\nCohSurvived}         {195\xspace}
\newcommand{\nCohSurvival}         {70.9\%\xspace}
\newcommand{\nCohExit}             {80\xspace}
\newcommand{\nCohMcapOne}          {\$18.05 billion\xspace}
\newcommand{\nCohMcapTwo}          {\$11.48 billion\xspace}
\newcommand{\nCohChange}           {36.4\%\xspace}
\newcommand{\nCohSurvivorChange}   {35.5\%\xspace}
\newcommand{\nCohExitValue}        {\$236 million\xspace}
\newcommand{\nCohGiniOne}          {0.887\xspace}
\newcommand{\nCohGiniTwo}          {0.924\xspace}
\newcommand{\nCohTopDecileOne}     {80.1\%\xspace}
\newcommand{\nCohTopDecileTwo}     {90.3\%\xspace}
\newcommand{\nCohBottomShare}      {19.9\%\xspace}
\newcommand{\nCohMedianOne}        {\$5.2 million\xspace}

\newcommand{\nPanCount}                {133\xspace}
\newcommand{\nPanCapitalised}          {77\xspace}

\newcommand{\nPanKept}                 {65\xspace}
\newcommand{\nPanKeptShare}            {84\%\xspace}
\newcommand{\nPanLeft}                 {6\xspace}
\newcommand{\nPanLeftShare}            {8\%\xspace}
\newcommand{\nPanGone}                 {6\xspace}
\newcommand{\nPanGoneShare}            {8\%\xspace}
\newcommand{\nPanStillTradedShare}     {92\%\xspace}

\newcommand{\nPanKeptUncapitalised}    {six\xspace}
\newcommand{\nPanKeptMcapOne}          {\$6.17 billion\xspace}
\newcommand{\nPanKeptMcapTwo}          {\$4.69 billion\xspace}
\newcommand{\nPanKeptRetained}         {76\%\xspace}
\newcommand{\nPanKeptMedian}           {21.8\%\xspace}
\newcommand{\nPanGrew}                 {six\xspace}
\newcommand{\nPanLostNinety}           {seventeen\xspace}
\newcommand{\nPanLostNinetyNine}       {four\xspace}
\newcommand{\nPanKeptPriced}           {59\xspace}
\newcommand{\nPanTopOneName}           {Theoriq\xspace}
\newcommand{\nPanTopOne}               {12.78$\times$\xspace}
\newcommand{\nPanTopTwoName}           {Venice.ai\xspace}
\newcommand{\nPanTopTwo}               {12.51$\times$\xspace}
\newcommand{\nPanHoldingOneName}       {NEAR Protocol\xspace}
\newcommand{\nPanHoldingOneRetained}   {1.03$\times$\xspace}
\newcommand{\nPanHoldingTwoName}       {Venice.ai\xspace}
\newcommand{\nPanHoldingTwoRetained}   {12.51$\times$\xspace}
\newcommand{\nPanHoldingsShare}        {70.5\%\xspace}
\newcommand{\nPanExOutlier}            {38\%\xspace}
\newcommand{\nPanCoded}                {75\xspace}
\newcommand{\nPanCodedMcap}            {\$6.92 billion\xspace}
\newcommand{\nPanIncubCount}           {26\xspace}

\newcommand{\nPanInfraCount}           {27\xspace}

\newcommand{\nPanInfraMean}            {\$174 million\xspace}
\newcommand{\nPanInfraShare}           {67.8\%\xspace}
\newcommand{\nPanFinCount}             {17\xspace}

\newcommand{\nPanFinMean}              {\$66.7 million\xspace}
\newcommand{\nPanFinShare}             {16.4\%\xspace}
\newcommand{\nPanCreativeCount}        {20\xspace}

\newcommand{\nPanFinCountShare}        {18.9\%\xspace}
\newcommand{\nPanMedianTrust}          {53\%\xspace}
\newcommand{\nPanAliveTrust}           {100\%\xspace}
\newcommand{\nPanMedianLaunchpad}      {29\%\xspace}

\newcommand{\nPanMedianDeFAI}          {9\%\xspace}

\newcommand{\nPanMedianContent}        {20\%\xspace}

\newcommand{\nPanAliveProtocol}        {76\%\xspace}

\newcommand{\nPanAliveGame}            {27\%\xspace}

\newcommand{\nRepCount}            {1{,}420\xspace}
\newcommand{\nRepToken}            {979\xspace}
\newcommand{\nRepRepo}             {805\xspace}
\newcommand{\nRepBoth}             {364\xspace}
\newcommand{\nRepTokenOnly}        {615\xspace}
\newcommand{\nRepRepoOnly}         {441\xspace}

\newcommand{\nRepRepoOnlyActive}   {266\xspace}
\newcommand{\nRepLiveBoth}         {120\xspace}
\newcommand{\nRepLiveNeither}      {87\xspace}
\newcommand{\nRepLiveCodeOnly}     {107\xspace}
\newcommand{\nRepLiveTokenOnly}    {50\xspace}
\newcommand{\nRepLiveShare}        {33\%\xspace}
\newcommand{\nRepPool}             {560\xspace}
\newcommand{\nRepGroups}           {456\xspace}
\newcommand{\nRepKwLegacy}         {82\xspace}
\newcommand{\nRepKwStandards}      {475\xspace}
\newcommand{\nRepKwBoth}           {three\xspace}
\newcommand{\nRepSingleShare}      {88\%\xspace}
\newcommand{\nRepLargestGroup}     {eight\xspace}

\newcommand{\nGovDescribed}                {750\xspace}
\newcommand{\nGovVocabulary}               {91\xspace}
\newcommand{\nGovWithAgent}                {44\xspace}
\newcommand{\nGovUtilityCount}             {23\xspace}

\newcommand{\nGovActorCount}               {10\xspace}
\newcommand{\nGovActorMcap}                {\$1.39 billion\xspace}
\newcommand{\nGovActorShare}               {9.0\%\xspace}
\newcommand{\nGovSubjectCount}             {11\xspace}

\newcommand{\nGovVocabularyShare}          {12.1\%\xspace}
\newcommand{\nGovActorLargest}             {Internet Computer\xspace}
\newcommand{\nGovActorMcapExLargest}       {\$50.0 million\xspace}

\newcommand{\nXFourSettlements}            {432{,}913\xspace}
\newcommand{\nXFourPeak}                   {240{,}098\xspace}
\newcommand{\nXFourValue}                  {\$130{,}171\xspace}
\newcommand{\nXFourPayers}                 {7{,}079\xspace}
\newcommand{\nXFourPayees}                 {1{,}341\xspace}
\newcommand{\nXFourAsset}                  {USDC\xspace}
\newcommand{\nXFourChain}                  {Base\xspace}
\newcommand{\nXFourContract}               {\texttt{SettlementRouter}\xspace}
\newcommand{\nXFourMedian}                 {\$0.0012\xspace}
\newcommand{\nXFourPNinety}                {\$0.0057\xspace}

\newcommand{\nXFourMean}                   {\$0.30\xspace}
\newcommand{\nXFourLargest}                {\$9{,}997\xspace}
\newcommand{\nXFourUnderCentShare}         {94.9\%\xspace}
\newcommand{\nXFourUnderDollarShare}       {99.2\%\xspace}

\newcommand{\nXFourPayeesHigh}             {47\xspace}
\newcommand{\nXFourPayersLow}              {29\xspace}
\newcommand{\nXFourPayersHigh}             {2{,}744\xspace}
\newcommand{\nXFourSecondMonth}            {May 2026\xspace}
\newcommand{\nXFourSecondValue}            {\$18.5k\xspace}
\newcommand{\nXFourSecondSettlements}      {20{,}576\xspace}
\newcommand{\nXFourBurstDate}              {4 December 2025\xspace}
\newcommand{\nXFourBurstSettlements}       {2{,}544\xspace}
\newcommand{\nXFourBurstValue}             {\$103{,}687\xspace}
\newcommand{\nXFourBurstAddresses}         {1{,}473\xspace}
\newcommand{\nXFourBurstRelays}            {622\xspace}
\newcommand{\nXFourBurstForwarded}         {\$43{,}586\xspace}
\newcommand{\nXFourBurstForwardedShare}    {42\%\xspace}
\newcommand{\nXFourBurstUniform}           {763\xspace}
\newcommand{\nXFourBurstUniformValue}      {\$100.01\xspace}
\newcommand{\nXFourBurstSeed}              {\$9{,}997\xspace}
\newcommand{\nXFourPeakMonth}              {March 2026\xspace}
\newcommand{\nXFourPeakValue}              {\$326\xspace}
\newcommand{\nXFourPeakMedian}             {\$0.0012\xspace}
\newcommand{\nXFourPeakLargest}            {\$0.10\xspace}
\newcommand{\nXFourDecMedian}              {\$8.63\xspace}
\newcommand{\nXFourDecLargest}             {\$9{,}997\xspace}
\newcommand{\nXFourDecMonth}               {December 2025\xspace}
\newcommand{\nXFourDecSettlements}         {3{,}198\xspace}
\newcommand{\nXFourDecValue}               {\$108.7k\xspace}
\newcommand{\nXFourFirstMonth}             {November 2025\xspace}
\newcommand{\nXFourSustainedFrom}          {February 2026\xspace}
\newcommand{\nXFourLastMonth}              {August 2026\xspace}
\newcommand{\nXFourLastSettlements}        {1{,}098\xspace}
\newcommand{\nXFourDeclineShare}           {0.5\%\xspace}
\newcommand{\nStdPool}                     {560\xspace}

\newcommand{\nStdAaRepos}                  {2\xspace}
\newcommand{\nErcChains}                   {16\xspace}
\newcommand{\nErcRegistrations}            {530{,}266\xspace}
\newcommand{\nErcRegistrants}              {321{,}100\xspace}
\newcommand{\nErcTopChain}                 {BNB Chain\xspace}

\newcommand{\nErcTail}                     {74{,}518\xspace}
\newcommand{\nErcTailChains}               {thirteen\xspace}
\newcommand{\nErcBNBRegistrations}         {324{,}450\xspace}
\newcommand{\nErcBNBRegistrants}           {264{,}877\xspace}
\newcommand{\nErcBNBRatio}                 {1.2\xspace}
\newcommand{\nErcBaseRegistrations}        {80{,}641\xspace}

\newcommand{\nErcEthereumRegistrations}    {50{,}657\xspace}

\newcommand{\nErcEthereumRatio}            {4.5\xspace}

\newcommand{\nErcCeloRegistrants}          {728\xspace}
\newcommand{\nErcCeloRatio}                {13.5\xspace}

\newcommand{\nErcMonadRegistrants}         {8{,}581\xspace}
\newcommand{\nErcMonadRatio}               {1.2\xspace}

\newcommand{\nErcAvalancheRegistrants}     {54\xspace}
\newcommand{\nErcAvalancheRatio}           {33.5\xspace}

\newcommand{\nErcRepCalls}                 {505{,}567\xspace}
\newcommand{\nErcRepSenders}               {21{,}872\xspace}
\newcommand{\nErcRepSenderShare}           {7\%\xspace}
\newcommand{\nErcRepBNBCalls}              {29{,}626\xspace}
\newcommand{\nErcRepBNBSenders}            {108\xspace}

\newcommand{\nErcRepBaseMultiple}          {5.7\xspace}

%% file: tex/01-introduction.tex
The rapid evolution of Web3 technologies has created expansive decentralized ecosystems spanning multiple blockchain networks, with Total Value Locked (TVL) exceeding \$100 billion across protocols~\cite{defillamaDefiLlama2025}. However, these ecosystems continue to face persistent challenges, including complex user interfaces, inefficient decision-making processes, and limited autonomous capabilities, which hinder their mainstream adoption~\cite{zhouSoKDecentralizedFinance2023a}. Concurrently, the advancement of large language models (LLMs) and AI agent technologies has enabled sophisticated reasoning and autonomous task execution. Nevertheless, such AI systems currently lack inherent mechanisms for trustless operation and decentralized coordination~\cite{zhengUnderstandingLargeLanguage2024a,karimAIAgentsMeet2025}.

Fortunately, the integration of Web3 and AI technologies can address mutual limitations: Web3 provides cryptographic security and decentralized infrastructure for AI agents, while AI agents enhance Web3 accessibility, efficiency, and intelligent automation~\cite{kayikciBlockchainMeetsMachine2024,huangOverviewWeb3Technology2024a,salehBlockchainSecureDecentralized2024}. Current Web3-AI Agent implementations demonstrate significant market traction and technical sophistication~\cite{kassisWeb3MeetsAI2025}. Projects range from simple chatbot interfaces that assist with blockchain queries to autonomous agents managing multi-million dollar DeFi portfolios, executing complex cross-chain transactions, and actively participating in decentralized governance~\cite{coinmarketcapTopAIBig}.

This paper presents the first comprehensive landscape analysis of the intersection between Web3 and AI agents, organized around five research questions (RQs) representing critical aspects of Web3-AI Agent integration. Firstly, we develop a taxonomy and map the existing market landscape (RQ1), revealing distinct patterns in both project distribution and capitalization. Building upon this landscape analysis, we explore four key integrations. We investigate how AI agents participate in and optimize decentralized finance (RQ2) and enhance Web3 governance mechanisms (RQ3). We examine how AI agents strengthen Web3 security through intelligent vulnerability detection and automated smart contract auditing (RQ4). Furthermore, we assess how Web3's native trust infrastructure creates robust reliability frameworks for AI agent operations (RQ5).

Through the analysis of \nPopCount Web3-AI agent tokens collectively valued at
\nPopMcap, \nClsCount of which we classify in depth, we establish
that the intersection of Web3 and AI technologies creates integrations
that address fundamental challenges in both domains while enabling
autonomous decentralized operations. Infrastructure projects dominate the
landscape in both project count (\nClsInfraCount of \nClsCount classified)
and capitalization (\nClsInfraShare), yet the ecosystem also hosts sizable
clusters of financial tooling and project incubation. Value is extremely
concentrated: the top decile of tokens holds \nPopTopDecileShare of
capitalization, and across the archived cohort aggregate value fell
\nCohChange over \nWindow even as the population multiplied.

The integration yields gains in both directions: AI agents acquire trustless execution guarantees from Web3 infrastructure, while Web3 systems gain accessibility and automation from AI. Significant challenges remain in scalability, security, ethics, and technical integration.

Our analysis offers the following key contributions:

\begin{itemize}
\item We develop a taxonomy organizing \nClsCount Web3 AI agent projects into four primary categories and ten subcategories (Figure~\ref{fig:landscape}), and extend the analysis with a second snapshot \nWindow later to create a two-timepoint longitudinal design.
\item We demonstrate specific mechanisms through which AI agents enhance Web3 operations across three critical domains: decentralized finance through autonomous strategy execution and intelligent portfolio optimization, governance through automated analysis and community engagement, and security through intelligent vulnerability detection and automated smart contract auditing.
\item We examine how Web3 trust mechanisms enable reliable AI agent operations through cryptographic guarantees, decentralized verification, and transparent accountability systems.
\item We provide open-source access to our full dataset at both timepoints,
including the second-round collection, the coding manual, and the per-project
classification with its provenance:
\url{https://github.com/shenyimings/Web3-AI-Agents}.
\end{itemize}

%% file: tex/02-background.tex
\subsection{Web3 Ecosystem}

Web3 represents the decentralized evolution of the internet, built on blockchain infrastructure where users maintain ownership of their digital assets and identities through cryptographic protocols and smart contracts~\cite{huangOverviewWeb3Technology2024a}. The ecosystem spans multiple blockchain networks, including Ethereum~\cite{buterinNextgenerationSmartContract2014}, Solana~\cite{solanaSolana2025}, and layer-2 solutions~\cite{coinbaseBase2025,optimismOptimism2025}, each supporting a diverse range of decentralized applications (DApps) that span decentralized finance (DeFi) protocols, gaming, and content creation platforms~\cite{zhengOverviewSmartContracts2020}. Currently, the Total Value Locked (TVL) across all Web3 protocols exceeds \$100 billion~\cite{defillamaDefiLlama2025}, demonstrating the significant scale and adoption of this emerging digital economy.

\subsection{Large Language Model (LLM)}

LLMs represent advanced AI systems trained on vast text corpora to understand and generate human-like text, enabling sophisticated reasoning and interaction capabilities~\cite{zhaoSurveyLargeLanguage2025}. Modern LLMs like GPT~\cite{openaiGPT4o2024}, Claude~\cite{Claude37Sonnet20250224}, and Gemini~\cite{googleGemini25Pro2025} demonstrate remarkable proficiency in natural language understanding, code generation, and complex problem-solving tasks that form the cognitive foundation for autonomous agent systems~\cite{wangSurveyLargeLanguage2024}.

\subsection{Web3-AI Agent}

\begin{figure}[h]
    \centering
    \includegraphics[width=\linewidth]{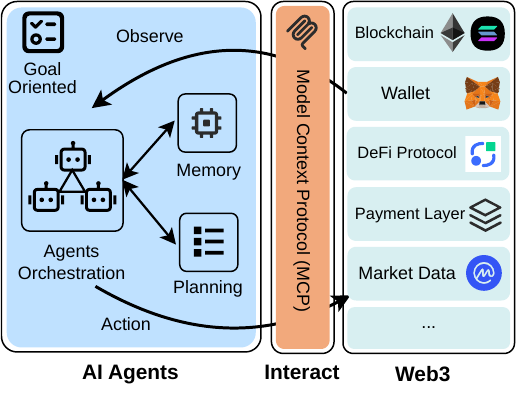}
    \caption{A Typical Architecture of Web3-AI Agents}
    \vspace{-1em}
    \label{fig:architecture}
\end{figure}

Figure \ref{fig:architecture} illustrates a typical Web3-AI Agent architecture where LLMs serve as the core reasoning engine, equipped with essential agent capabilities including planning, memory, and multi-agent orchestration. These agents interact with blockchain infrastructure through the Model Context Protocol (MCP)~\cite{anthropicModelContextProtocol2025}, a standardized interface for communication between LLMs and external tools or resources. This architecture enables AI agents to function as intelligent front-ends for Web3 interactions, handling complex tasks such as navigating multiple blockchains, managing gas fees, interfacing with various protocols and tokens, connecting to decentralized applications, and securing sensitive information like seed phrases.

Web3-AI Agent currently manifests in diverse forms, including autonomous trading bots~\cite{wayfinderHomeWayfinder2025}, DeFi portfolio managers~\cite{nexoNexoYourWealth2025}, cross-chain transaction assistants~\cite{heyanonHeyAnon2025}, and intelligent wallet interfaces~\cite{armorArmorWalletAI2025}. The ecosystem demonstrates rapid growth with projects ranging from simple chatbot interfaces for blockchain queries to sophisticated autonomous agents capable of executing complex multi-step DeFi strategies and managing entire crypto portfolios without human intervention~\cite{nartey2024decentralized}.
\subsection{Agent Interoperability Standards}
\label{subsec:standards}

Most Web3-AI agent projects integrate agents through bespoke, per-project
mechanisms. A standards layer has begun to form underneath them, covering
identity, payments, accounts and communication.

\paratitle{Agent identity and reputation} ERC-8004, ``Trustless
Agents''~\cite{erc8004TrustlessAgents2025}, defines three registries: an
Identity registry issuing an ERC-721 token per agent with a resolvable metadata
URI, a Reputation registry recording attestations about an agent, and a
Validation registry recording verifications of an agent's work. It reached
Ethereum mainnet on 29 January 2026 and is the first broadly deployed
on-chain standard for agent identity.

\paratitle{Agent payments} The x402 protocol~\cite{coinbaseX402Protocol2025}
revives the HTTP 402 \textit{Payment Required} status code, letting a server
answer a request with a machine-readable price that a client agent settles in
stablecoins before retrying. Introduced by Coinbase in May 2025, its governance
moved to the x402 Foundation under the Linux Foundation on 2 April 2026.
Google's Agent Payments Protocol (AP2)~\cite{googleAP2Protocol2025} addresses
the same problem from the authorization side, using verifiable-credential
mandates to bound what an agent may spend on a user's behalf, and Virtuals'
Agent Commerce Protocol standardizes agent-to-agent commercial negotiation.

\paratitle{Agent accounts} Account abstraction supplies the wallet these
payment protocols presuppose. ERC-4337~\cite{erc4337AccountAbstraction2021}
introduces programmable smart-contract accounts through a separate user
operation mempool, and EIP-7702, shipped in Ethereum's Pectra upgrade, allows an
externally owned account to temporarily adopt contract code, so an agent can act
under scoped, revocable authority rather than a bare private key.

\paratitle{Agent communication} Outside Web3, the Model Context Protocol
(MCP)~\cite{anthropicModelContextProtocol2025} standardizes an agent's access to
tools and data sources, and Agent2Agent
(A2A)~\cite{googleAnnouncingAgent2AgentProtocol2025} standardizes delegation
between agents. Neither is blockchain-specific, but both appear throughout the
Web3 agent codebases we survey, and ERC-8004 is explicitly designed to supply
the trust layer A2A leaves unspecified.

These standards are not merely proposals: Sections~\ref{subsec:x402}
and~\ref{subsec:sybil} measure their on-chain adoption directly.

\begin{figure*}
    \centering
    \includegraphics[width=\textwidth]{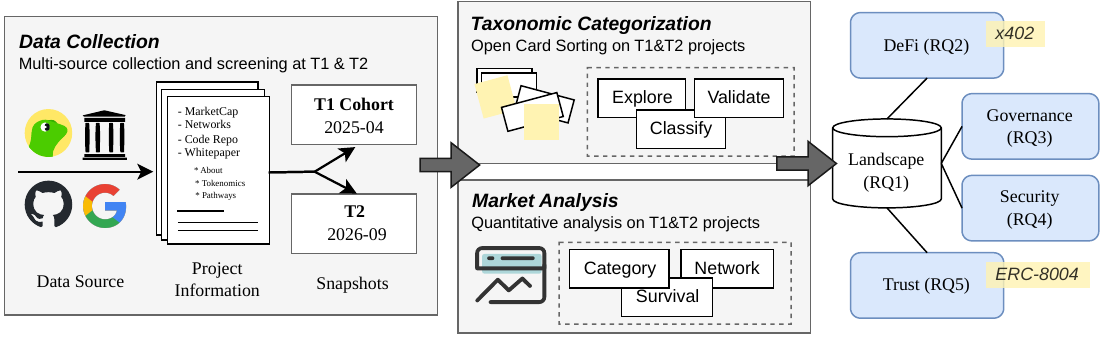}
    \caption{Research Methodology}
    \label{fig:methodology}
\end{figure*}

%% file: tex/03-methodology.tex
This study employs a mixed-methods approach combining systematic data collection, qualitative categorization, and quantitative market analysis to investigate the Web3-AI Agent landscape. The overall methodology is illustrated in Figure~\ref{fig:methodology}.

\subsection{Data Collection}\label{subsec:collection}

This study examines the Web3-AI agent ecosystem at two timepoints. We refer to \nDateOne as T1 and \nDateTwo as T2, and to the \nWindow between them as the \textit{study window}. At T2 we collect from two sources: CoinGecko~\cite{coingeckoCoinGecko2025} for market-listed tokens with verifiable trading, and GitHub~\cite{githubinc.Github2025} for open-source projects, including early-stage projects without a token. At T1 we reconstruct the ecosystem from archived CoinGecko category pages and date-restricted search, which lets us measure survival and value change (Section~\ref{subsec:baselines}). Our data sources, keywords, and datasets are available in our online repository.

\subsubsection{Keyword Set}

We compiled keywords in two groups. The first covers ecosystem-level terminology: ``Web3 AI'', ``blockchain AI'', ``Web3 agent'', ``Crypto agent'', ``DeFi agent'', ``Web3 AI Agent'', ``Crypto AI Agent'', ``Crypto LLM Agent'', ``Crypto LLM'', ``Crypto AI'', ``Blockchain Agent'', ``Chain Agent'', ``Decentralized AI Agent'', ``Blockchain LLM'', and ``DeFi AI''. We expanded these from an initial five terms using snowball sampling~\cite{goodmanSnowballSampling1961a}. The second group covers standards that emerged in 2025--2026: ERC-8004, x402, AP2, A2A, MCP, ERC-4337, EIP-7702, and related terms (20 total).
We applied each group separately to GitHub repositories created after April 2025. The two groups are nearly disjoint: \nRepKwLegacy repositories match only the ecosystem set, \nRepKwStandards only the standards set, and \nRepKwBoth match both.

\subsubsection{Token Collection}

We retrieve from CoinGecko every token listed under any of seven AI categories (\texttt{ai-agents}, \texttt{ai-agent-launchpad}, \texttt{ai-applications}, \texttt{ai-framework}, \texttt{ai-meme-coins}, \texttt{defai}, \texttt{artificial-intelligence}), totaling \nScrFrame listings, and read each token's capitalization at the cutoff date. A secondary pass retrieves the project description, homepage, repository links, and platform listings. 

Five screening rules govern the population, and none may reference capitalization rank or value share.
\textit{S1} requires that a token trade at the cutoff date, removing \nScrNotCapitalised non-capitalized listings. \textit{S2} removes claims on other tokens in the population, specifically staking or wrapped derivatives (\nScrDerivative listings worth \nScrDerivativeValue) and secondary listings (\nScrDuplicate listings from \nScrDuplicateGroups projects with multiple tokens), so that each project is counted once. \textit{S3} requires a self-description from CoinGecko, the project website, its whitepaper, or its repository. \textit{S5} excludes listings whose only category among the seven is \texttt{ai-meme-coins}, \nScrMeme listings worth \nScrMemeValue, on the grounds that meme listings are not projects with products. This rule turns on the category under which a listing enters the frame, not on CoinGecko's token-level meme label, which would otherwise exclude several widely cited agent projects. \textit{S4} requires AI or agent content in the description. We apply it when reading descriptions during coding rather than at screening, so its removals are reported with the coding in Section~\ref{subsec:classification}.

After \textit{S1}, \textit{S2}, and \textit{S5}, the frame holds \nScrMechanical tokens worth \nScrMechanicalMcap. Applying \textit{S3} leaves the study population of \nPopCount tokens worth \nPopMcap. The \nScrNoDescription listings with no description from any source are worth \nScrNoDescriptionValue, representing \nScrNoDescriptionShare of the screened value, and are recorded rather than discarded silently.

\subsubsection{Repository Collection}

From GitHub we collect repositories matching the keyword set and created after April 2025, \nRepPool in total. Repositories are resolved to projects by owner, with the project homepage domain as a secondary signal, giving \nRepGroups distinct projects; \nRepSingleShare of the resulting groups contain a single repository and the largest contains \nRepLargestGroup, so the many-to-one relation does not materially affect counts. Repository counts are not product counts (several groups are vendor sample code or demonstration portfolios), and standards adoption is accordingly reported per repository, never folded into the project taxonomy. We use GitHub for \textit{discovery} (restricted to repositories created after April 2025) and for \textit{measurement} (without that restriction).

\subsubsection{T1 Collection}\label{subsec:baselines}

We reconstruct T1 membership from Internet Archive Wayback Machine captures of CoinGecko's AI category pages near T1\footnote{The Wayback Machine captures pages at irregular intervals. We select \nDateOne (T1) as it was the only capture available in our target study window.}: \texttt{artificial-intelligence}, \texttt{ai-agents}, \texttt{ai-agent-launchpad}, and \texttt{defai}. We add further projects through Google searches that join the keywords of Section~\ref{subsec:collection} with OR and bound them with the \texttt{after:} and \texttt{before:} date operators, using queries of the form
\begin{center}
\small\texttt{"\textless KEYWORD\textgreater" OR "\textless KEYWORD\textgreater" OR \ldots{} after:2024-10-06 before:2025-04-06}
\end{center}
All candidates are screened with the rules above. We read each member's capitalization from CoinGecko at both T1 and T2. Because membership is fixed at T1, survival and value change are measured on a closed set rather than by comparing two open cross-sections.

\begin{figure*}
    \centering
    \includegraphics[width=\textwidth]{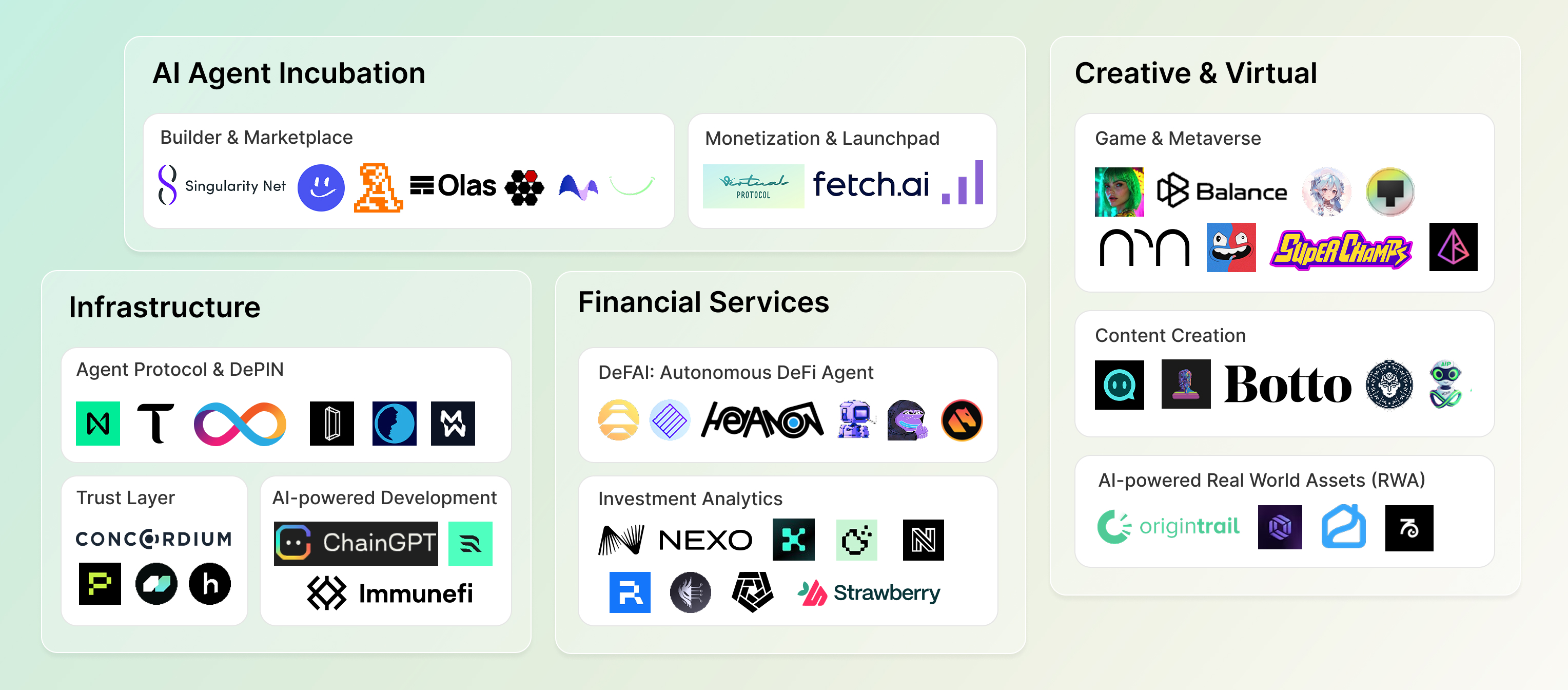}
    \caption{The Ecosystem Landscape of Web3-AI Agent Projects}
    \label{fig:landscape}
\end{figure*}

\subsection{Taxonomic Classification}\label{subsec:classification}

We develop a taxonomy through structured qualitative analysis using open card sorting adapted from established software engineering research practices~\cite{wohlinEmpiricalResearchMethods2003,wanSmartContractSecurity2021}.

\subsubsection{Classification Process}

Every token in the study population is coded against four primary categories and ten subcategories under a written coding manual that restates each subcategory as a discriminating test, released with the artifact. Coding proceeds in two layers: \nClsRule projects are assigned from CoinGecko's functional category labels where a label names a function rather than a theme and maps to exactly one subcategory, and the remaining \nClsRead are read individually. Two researchers code independently ($\kappa = 0.87$) and a third adjudicates disagreements. Every assignment carries its provenance and the evidence sentence.

Two residual outcomes are excluded from all per-project results: \nClsOutOfScope projects whose descriptions carry neither AI nor agent content, which is rule \textit{S4}, and \nClsUnclassifiable for which no description supports a functional reading. This leaves \nClsCount classified projects, which are the unit of every taxonomic, network and per-project valuation result reported in Section~\ref{sec:rq1}. Aggregate and concentration results are reported over the full \nPopCount-token population, which requires no classification.

Within \textit{Agent Protocol \& DePIN}, each project additionally carries a facet, \texttt{chain}, \texttt{compute}, \texttt{model} or \texttt{data}, recording what the project supplies. The facet is a strict refinement of the subcategory rather than a category of its own: cross-timepoint comparison uses the subcategory label.

\subsubsection{Taxonomy Framework}

\input{tables/taxonomy}

\footnotetext{\nClsMulti projects are classified under multiple subcategories due to
their multifaceted nature, and their full market capitalization is attributed
to each category they belong to in this analysis.}

This process produces a taxonomy comprising four primary categories and ten subcategories, as detailed in Table~\ref{tab:consolidated_taxonomy}:

\begin{itemize}
    \item \textbf{AI Agent Incubation} (\nClsIncubCount projects): Platforms for creating, deploying, and monetizing AI agents within Web3 environments. This category includes \textit{Builder \& Marketplace platforms} (\nClsBuilderCount projects) offering development tools and agent distribution systems, and \textit{Monetization \& Launchpad services} (\nClsLaunchpadCount projects) facilitating agent tokenization and funding mechanisms.
    \item \textbf{Infrastructure} (\nClsInfraCount projects): Foundational technologies enabling decentralized AI agent operations. Subcategories include \textit{Agent Protocol \& DePIN systems} (\nClsProtocolCount projects) providing core networking and computational resources, \textit{Trust Layer protocols} (\nClsTrustCount projects) establishing verification and reputation mechanisms, and \textit{AI-powered Development Tools} (\nClsCopilotCount projects) offering specialized programming environments and security tooling for agent creation.
    \item \textbf{Financial Services} (\nClsFinCount projects): Applications leveraging AI agents for decentralized finance operations and market analysis. This encompasses \textit{DeFAI Agents} (\nClsDeFAICount projects) performing autonomous trading and portfolio management, and \textit{Investment Analytics tools} (\nClsAnalyticsCount projects) providing market intelligence and risk assessment capabilities.
    \item \textbf{Creative \& Virtual} (\nClsCreativeCount projects): Projects utilizing AI agents in gaming, content creation, and asset management contexts. Subcategories include Game \& Metaverse applications (\nClsGameCount projects), Content Creation platforms (\nClsContentCount projects), and AI-powered Real World Asset (RWA) management systems (\nClsRWACount projects).
\end{itemize}

\subsection{Market Analysis}

We conduct quantitative market analysis using market capitalization data to understand the economic significance and distribution patterns across our taxonomy. Market capitalization data provides insights into investor confidence and project maturity within each category~\cite{huangOverviewWeb3Technology2024a}. Cross-sectional valuations are CoinGecko market capitalizations as of \nDateTwo; T1 values are read as described in Section~\ref{subsec:baselines}.

\subsection{Cross-Timepoint Measurement}\label{subsec:t2}

\paratitle{Monthly price series} Figure~\ref{fig:mcap_trend} additionally draws on one \texttt{market\_chart/range} call per classified project from \nTrendFirst, and every snapshot statistic is dated \nDateTwo.

\paratitle{Cohort matching} To trace the \nPanCount T1 projects found through search to the 2026 collection, we match against the complete CoinGecko coin listing, maintain an explicit map of renames and fall back to the token contract address, which does not drift when a name does. ``Left the category'', ``no longer traded'' and ``still an AI agent'' are reported as three distinct outcomes throughout.

%% file: tables/taxonomy.tex
\begin{table*}[ht]
\centering
\caption{Consolidated Web3-AI Agent Taxonomy with Market Metrics\protect\footnotemark}
\label{tab:consolidated_taxonomy}
\resizebox{\textwidth}{!}{%
\begin{tabular}{llcll}
\toprule
\textbf{Primary Category} & \textbf{Avg. MCap (\$M)} & \textbf{Subcategory} & \textbf{Count} & \textbf{Representative Projects} \\
\midrule
\multirow{2}{*}{AI Agent Incubation (213)} & \multirow{2}{*}{11.5} & Builder \& Marketplace & 160 & Unibase, EigenCloud (prev. EigenLayer), MultiversX \\
\cline{3-5}
 &  & Monetization \& Launchpad & 69 & Artificial Superintelligence Alliance, AWE Network, tokenbot \\
\cline{1-5}
\multirow{3}{*}{Infrastructure (305)} & \multirow{3}{*}{38.5} & Agent Protocol \& DePIN & 226 & NEAR Protocol, Bittensor, Internet Computer \\
\cline{3-5}
 &  & Trust Layer & 70 & Concordium, Sapien, RedTeam \\
\cline{3-5}
 &  & AI-powered Development & 27 & TokenOS AI, Gittensor, XIIID \\
\cline{1-5}
\multirow{2}{*}{Financial Services (278)} & \multirow{2}{*}{4.6} & DeFAI Agents & 185 & Ribbita by Virtuals, Derive, Velvet \\
\cline{3-5}
 &  & Investment Analytics & 107 & Arkham, KAITO, EfficientFrontier \\
\cline{1-5}
\multirow{3}{*}{Creative \& Virtual (137)} & \multirow{3}{*}{4.3} & Game \& Metaverse & 64 & MetYa, Victoria VR, OKZOO \\
\cline{3-5}
 &  & Content Creation & 68 & Bitcast, 404---GEN, Everlyn \\
\cline{3-5}
 &  & AI-powered RWA & 8 & PinLink, Zipcode, GAIB \\
\bottomrule
\end{tabular}%
}
\end{table*}

%% file: tex/04-rq1.tex
This section analyzes the Web3-AI Agent token population using two distinct frames.
The screened population of \nPopCount capitalized tokens supports aggregate and
concentration results. Per-project attributes (category, blockchain, valuation)
are reported over the \nClsCount projects whose descriptions support functional
classification (Section~\ref{subsec:classification}). This split is close to
exhaustive: classification fails only when a description carries no AI or agent
content. The classified projects hold \nClsMcap of the population's \nPopMcap, so
per-project results describe essentially the entire capitalized ecosystem, and the
residual is characterized by the concentration results in
Section~\ref{subsec:longitudinal}. These patterns frame the four integration
domains discussed in the subsequent research questions.

\subsection{Dilution Without Growth}
\label{subsec:longitudinal}

We compare the T2 population against the T1 cohort (Section~\ref{subsec:baselines}). The result inverts the growth narrative that dominated
the sector in 2025.

We compare a closed cohort because the two open cross-sections cannot be directly
differenced: the frame widened between them. Of the \nCohCount tokens capitalized
at T1, \nCohSurvived still trade at T2, a survival rate of \nCohSurvival. Their
aggregate capitalization fell from \nCohMcapOne to \nCohMcapTwo, a decline of
\nCohChange. Almost none of this loss comes from projects disappearing: the
\nCohExit members that stopped trading took \nCohExitValue with them, while the
survivors alone lost \nCohSurvivorChange of their value. The ecosystem did not
shrink by attrition; it repriced.

Issuance moved in the opposite direction. The screened population at T2 holds
\nPopCount tokens against a T1 cohort of \nCohCount. The widened frame explains
part of this gap, but not the price: the median token is now worth
\nPopMedian against \nCohMedianOne at T1. New tokens arrived faster than capital,
and the new arrivals were small.

\begin{figure}[t]
    \centering
    \includegraphics[width=\linewidth]{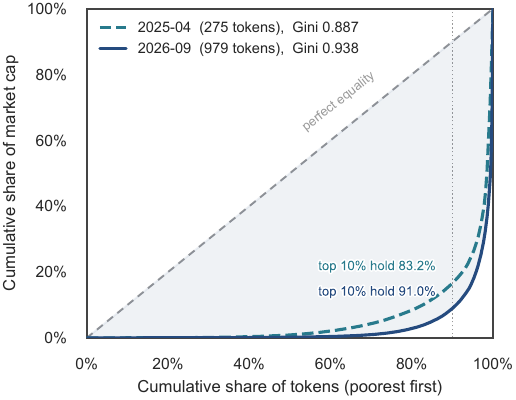}
    \caption{Lorenz curves of market capitalization at both timepoints. The
    2026 curve lies almost on the axes: capitalization is now so concentrated
    that the poorest 90\% of tokens hold \nPopBottomShare of the value, against
    \nCohBottomShare for the T1 cohort.}
    \label{fig:mcap_lorenz}
\end{figure}

Concentration intensified rather than dispersed. The Gini coefficient of market
capitalization reaches \nPopGini across the population; the top 1\% of tokens
(\nPopTopPctCount projects) hold \nPopTopPctShare of all value and the top 5\%
hold \nPopTopTwentiethShare. Within the cohort, where the comparison is like for
like, the Gini rises from \nCohGiniOne to \nCohGiniTwo and the top-decile share
from \nCohTopDecileOne to \nCohTopDecileTwo. The lower tail dominates in count
but is negligible in value:
\nPopSubHundredKShare of tokens are capitalized below \$100k and a further
\nPopHundredKToMShare below \$1 million, while only \nPopAboveHundredMCount
tokens (\nPopAboveHundredMShare) exceed \$100 million
(Figure~\ref{fig:mcap_lorenz}). Individual trajectories follow the aggregate.
The Artificial Superintelligence Alliance (formerly Fetch.ai), the largest agent
project at T1, retains only a fraction of its former capitalization
(Section~\ref{subsec:casestudy}).

\begin{figure}[t]
    \centering
    \includegraphics[width=\linewidth]{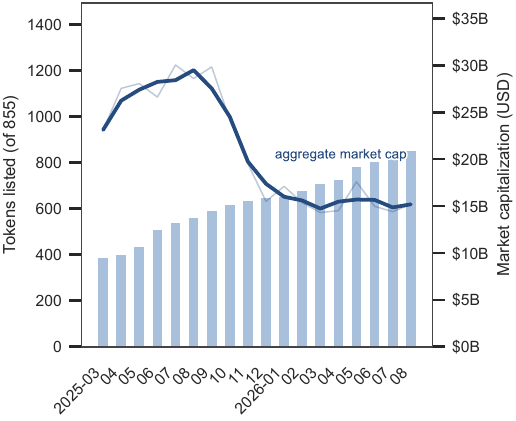}
    \caption{Listings and capitalization move in opposite directions over the
    \nTrendSpan from \nTrendFirst to \nTrendLast, the last full month before the
    second cutoff. Bars count how many of the \nClsCount classified projects
    had a live listing in each month, rising from \nTrendLiveOne to
    \nTrendLiveTwo. The line shows their aggregate capitalization, falling from
    \nTrendMcapOne to \nTrendMcapTwo, drawn faint for monthly observations and
    bold for a centred three-month mean.}
    \label{fig:mcap_trend}
\end{figure}

\observation{Over the \nWindow study window the closed T1 cohort lost \nCohChange of its capitalization while \nCohSurvival of its members continued to trade, and the token count multiplied around it: the ecosystem diluted rather than grew, and value concentrated further into a shrinking head.}

The longitudinal comparison changes what concentration means. What looked at T1
like an early-stage ecosystem awaiting broader capital allocation proved to be a
structure where entry is cheap, abundant, and almost entirely uncompensated. The
modal Web3-AI agent token is not early-stage but dormant. Single-snapshot designs
cannot separate these readings, which is the main methodological argument for the
two-timepoint frame adopted here.

\subsection{Distribution and Market Patterns}

\subsubsection{Taxonomic Distribution}

Projects distribute across the taxonomy far more evenly than capital does.
\textit{Infrastructure} is the largest primary category, with \nClsInfraCount
of the \nClsCount classified projects: the chains, compute networks, model
markets and data layers on which agents run. \textit{Financial Services}
follows with \nClsFinCount, \textit{AI Agent Incubation} with
\nClsIncubCount, and \textit{Creative \& Virtual} with \nClsCreativeCount.
A project assigned to multiple categories appears in each count, so the four
figures sum to more than \nClsCount.

\begin{figure}[t]
    \centering
    \includegraphics[width=\linewidth]{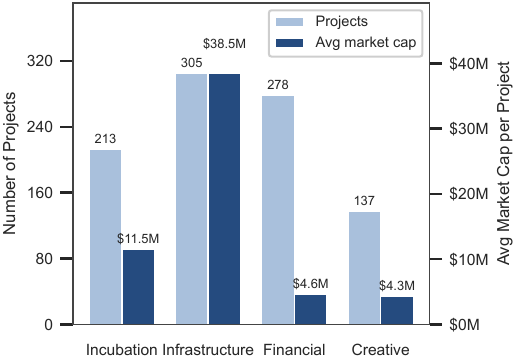}
    \caption{Project count and average market capitalization by primary
    category across the \nClsCount classified projects. Counts are category
    memberships: a project assigned to two categories is counted in both.}
    \label{fig:marketcap}
\end{figure}

Capitalization distributes very differently (Figure~\ref{fig:marketcap}). Of
the \nClsMcap held by classified projects, \textit{Infrastructure} accounts
for \nClsInfraMcap, or \nClsInfraShare, at \nClsInfraMean per project. The
other three categories together hold about a quarter: \textit{AI Agent
Incubation} \nClsIncubMcap (\nClsIncubShare, \nClsIncubMean per project),
\textit{Financial Services} \nClsFinMcap (\nClsFinShare, \nClsFinMean) and
\textit{Creative \& Virtual} \nClsCreativeMcap (\nClsCreativeShare,
\nClsCreativeMean). A multi-category project's capitalization is attributed
to each of its categories, so shares sum to more than 100\%. The foundational
layer's per-project valuation is four to nine times that of the application
layers, which are populous but lightly capitalized.

Within \textit{Infrastructure}, the \nClsFacetChainCount chain projects hold
\nClsFacetChainMcap, or \nClsFacetChainShare of the subcategory, at
\nClsFacetChainMean per project: the highest average in the taxonomy. Model
networks follow at \nClsFacetModelMean per project across \nClsFacetModelCount
projects, compute at \nClsFacetComputeMean across \nClsFacetComputeCount, and
data at \nClsFacetDataMean across \nClsFacetDataCount. Scarcity, not volume,
commands value here: the facets with the fewest projects achieve the highest
per-project valuations.

\observation{\textit{Infrastructure} holds \nClsInfraShare of classified capitalization on \nClsInfraCount projects, and the \nClsFacetChainCount chain projects within it average \nClsFacetChainMean each: project counts spread across the taxonomy, but value concentrates almost entirely in the foundational layer.}

At subcategory level, the distribution is even more uneven
(Table~\ref{tab:consolidated_taxonomy}). \textit{Agent Protocol \& DePIN} is
both the largest subcategory at \nClsProtocolCount projects and the richest at
\nClsProtocolMean per project. \textit{DeFAI Agents} follows in count with
\nClsDeFAICount projects but averages only \nClsDeFAIMean; \textit{Builder \&
Marketplace} holds \nClsBuilderCount projects worth \nClsBuilderMcap in total.
\textit{Monetization \& Launchpad} is smaller at \nClsLaunchpadCount projects
but averages \nClsLaunchpadMean, second only to the protocol layer among the
larger subcategories. At the other end, \textit{AI-powered Development}
comprises \nClsCopilotCount projects worth \nClsCopilotMcap in total, or
\nClsCopilotMean each: development tooling for agent builders is a real
activity in this ecosystem, but one conducted largely in public repositories
rather than financed through a token (Section~\ref{subsec:coverage}).

\subsubsection{Cross-Category Integration}

Functional boundaries are porous. \nClsMulti of the \nClsCount projects
(\nClsMultiShare) carry more than one subcategory label, holding
\nClsMultiMcap, or \nClsMultiMcapShare of classified capitalization. The
frequent combinations are structural rather than arbitrary:
\nClsPairOneName (\nClsPairOne projects), \nClsPairTwoName (\nClsPairTwo),
\nClsPairThreeName (\nClsPairThree), and \nClsPairFourName (\nClsPairFour).
Each pairs a place to build or run agents with what makes them trustworthy,
deployable or payable: one integrated-platform strategy, expressed several ways.

\begin{figure}[t]
    \centering
    \includegraphics[width=\linewidth]{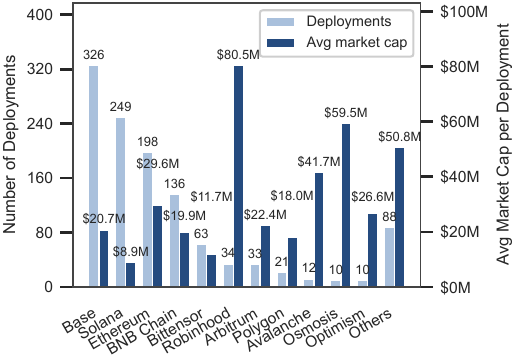}
    \caption{Deployment count and average market capitalization by blockchain
    network. A project deployed on several chains is counted on each, so the
    bars sum to deployments rather than to projects; networks hosting fewer
    than ten deployments are grouped as \textit{Others}.}
    \label{fig:network}
\end{figure}

\subsubsection{Blockchain Network}

Deployment is multi-chain by default (Figure~\ref{fig:network}).
\nClsListed of the \nClsCount classified projects (\nClsListedShare) list at
least one contract platform, with \nClsDeployments deployments across
\nClsChains networks; \nClsMultichain (\nClsMultichainShare) deploy on more
than one. Base hosts the most, \nClsChainBaseCount deployments
(\nClsChainBaseCountShare) worth \nClsChainBaseMcap in aggregate and
\nClsChainBaseMean per deployment. Solana ranks second in count with
\nClsChainSolanaCount (\nClsChainSolanaCountShare) but holds only
\nClsChainSolanaMcap, an average of \nClsChainSolanaMean. Ethereum hosts fewer,
\nClsChainEthereumCount (\nClsChainEthereumCountShare), worth
\nClsChainEthereumMcap at \nClsChainEthereumMean each, and BNB Chain
\nClsChainBNBCount (\nClsChainBNBCountShare) worth \nClsChainBNBMcap at
\nClsChainBNBMean. Bittensor (\nClsChainBittensorCount) and Robinhood Chain
(\nClsChainRobinhoodCount) follow, then Arbitrum (\nClsChainArbitrumCount) and
Polygon (\nClsChainPolygonCount), with the remaining networks hosting a few
deployments each. Robinhood Chain reports the highest average of any network,
\nClsChainRobinhoodMean, because a few large tokens are cross-listed there
rather than because agent projects build on it.

\observation{Web3-AI agent tokens deploy predominantly on low-fee execution layers, with Base carrying \nClsChainBaseCountShare of all deployments and Solana \nClsChainSolanaCountShare, but among the four networks that host most deployments the highest value per deployment is on Ethereum (\nClsChainEthereumMean): deployment and value follow different logics.}

\subsection{What a Market-Capitalization Frame Cannot See}
\label{subsec:coverage}

Both analyses above measure the ecosystem through its tokens. A parallel GitHub
collection lets us locate the boundary of that frame. We assembled a second,
independent frame from GitHub: repositories created between the two cutoffs
matching an extended keyword set (Section~\ref{subsec:collection}), resolved
from \nRepPool repositories to \nRepGroups distinct projects. Some of those also
carry a token, so the GitHub frame contributes \nRepRepoOnly projects the token
frame cannot see; the union is \nRepCount projects, of which \nRepToken carry a
token and \nRepRepo carry a repository.

\begin{table}[t]
\centering
\caption{Evidence available per project across the two sampling frames
(\nRepCount projects). The empty cell is structurally empty rather than zero: a project
with neither a token nor a public repository is invisible to both frames, so
both distributions are conditional on discovery.}
\label{tab:coverage}
\begin{tabular}{lcc}
\toprule
 & \textbf{Repository} & \textbf{No repository} \\
\midrule
\textbf{Token}    & \nRepBoth & \nRepTokenOnly \\
\textbf{No token} & \nRepRepoOnly & --- \\
\bottomrule
\end{tabular}
\end{table}

Table~\ref{tab:coverage} shows the result. Only \nRepBoth projects carry both
forms of evidence; \nRepTokenOnly are observable only as tokens and
\nRepRepoOnly only as repositories.\footnote{The low overlap is structural
rather than a matching failure: GitHub discovery is restricted to repositories
created after April 2025, while capitalized token projects overwhelmingly
predate that window.} Neither frame is close to a superset of the other. Of the
\nRepRepoOnly projects with no token at all, \nRepRepoOnlyActive have been
pushed to within \nStaleDays days of the cutoff and are under active
development. \textit{A large share of the live development activity in this
ecosystem is uncapitalized, and no token-based frame can observe it.}

For the \nRepBoth projects visible to both frames, we can ask what neither frame
answers alone: whether a project is alive. Using a market capitalization above
\nAliveFloor and a push within \nStaleDays days as the two liveness criteria,
\nRepLiveBoth projects are alive on both axes, \nRepLiveNeither on neither,
\nRepLiveCodeOnly show active development behind a token that has effectively
died, and \nRepLiveTokenOnly show a live token behind stale code. Only
\nRepLiveShare of dual-evidence projects are alive on both axes. The two
off-diagonal cells are both substantial and asymmetric: they are different
failure modes, and a market-capitalization-only study cannot distinguish either
from outright death.

\observation{Token capitalization and repository activity capture largely disjoint populations: of \nRepCount projects, \nRepRepoOnly have no token at all and \nRepRepoOnlyActive of those are actively developed, while among the \nRepBoth visible to both frames only \nRepLiveShare are alive on both axes.}

We therefore report liveness, which both frames can express, rather than forcing
stars and market capitalization onto a common scale. This also bounds the
token-based results: the T1 cohort and the \nPopCount T2 tokens were drawn from
aggregators, so they describe the capitalized ecosystem specifically, not the
Web3-AI agent ecosystem as a whole.

\subsection{Entry Moved Up the Stack; Value Did Not}
\label{subsec:shift}

Recoding T1 projects under the T2 manual enables a comparison the raw snapshots
do not support: how composition changed rather than how large each layer
was.\footnote{The two collections differ in breadth, so only shares are
compared, never levels. The coding is identical at both ends (the same manual,
the same coders, and a T1 label takes precedence wherever a project recurs), so
a change in composition cannot be an artifact of the coding.} At T1, the
\nPanCoded classified projects worth \nPanCodedMcap were spread evenly in count
across the taxonomy: \nPanInfraCount in \textit{Infrastructure},
\nPanIncubCount in \textit{AI Agent Incubation}, \nPanCreativeCount in
\textit{Creative \& Virtual}, and \nPanFinCount in \textit{Financial Services}.
By T2 this distribution has shifted. \textit{Financial Services} grew from
\nPanFinCount projects to \nClsFinCount, \textit{AI Agent Incubation} from
\nPanIncubCount to \nClsIncubCount, and \textit{Infrastructure} from
\nPanInfraCount to \nClsInfraCount. All categories multiplied in absolute
terms, but Financial Services expanded fastest, from \nPanFinCountShare of
assignments to \nClsFinCountShare.

Capitalization moved the other way. \textit{Infrastructure} held
\nPanInfraShare of capitalization at T1 and \nClsInfraShare at T2, while
\textit{Financial Services} fell from \nPanFinShare to \nClsFinShare despite
growing sixteen-fold in project count. Per-project value declined in every
category, with \textit{Infrastructure} dropping from \nPanInfraMean to
\nClsInfraMean and \textit{Financial Services} from \nPanFinMean to
\nClsFinMean, and the steepest declines came where entry was easiest. Building
an agent application became cheap; building the layer it runs on did not, and
the market still prices that difference.

Repository activity shows the same split. Among the \nActCount classified
projects with repository evidence, Infrastructure projects are the most active:
\nActInfraRate have been pushed to within \nStaleDays days of the cutoff,
against \nActIncubRate for Incubation, \nActFinRate for Financial Services, and
\nActCreativeRate for Creative \& Virtual. Capital and code agree about the
foundational layer and disagree elsewhere: the categories that grew fastest in
count are not the ones whose repositories are most reliably alive. These rates
are an upper bound, since a project that publishes a repository at all is more
likely to maintain it, and they compare categories rather than measure the
ecosystem's absolute activity.

\observation{Between the two snapshots \textit{Financial Services} grew disproportionately, from \nPanFinCount projects to \nClsFinCount, while capitalization concentrated further in the foundational layer, from \nPanInfraShare to \nClsInfraShare, and per-project value fell in every category.}

The landscape shows expanding multi-chain deployment, rapid entry at the
application layers, and value distributed far more unequally than project
count. The remaining research questions examine the four integration domains
this landscape supports: decentralized finance (RQ2), governance (RQ3),
security (RQ4), and trust infrastructure (RQ5).

\subsection{Case Study: T1 Projects at the Second Cutoff}
\label{subsec:casestudy}

The re-collection analyzed above describes the ecosystem as it stands today, so
projects that have since disappeared are absent from it by construction. To see
what became of the projects that were prominent at T1, we take the \nPanCount
projects that date-restricted search identifies at \nDateOne as a closed set and
follow them to the second cutoff, resolving each against the full CoinGecko
listing by name, by documented rename, and where names had drifted by token
contract address.

Survival is high; value retention is not. Of the \nPanCapitalised projects that
carried a market capitalization in 2025, \nPanKept (\nPanKeptShare) remain
listed under an AI-agent category, \nPanLeft (\nPanLeftShare) still trade
outside it, and \nPanGone (\nPanGoneShare) are no longer resolvable to a traded
token. On the narrowest reading \nPanKeptShare survived; on the broadest,
\nPanStillTradedShare still trade somewhere.

\begin{figure}[t]
    \centering
    \includegraphics[width=\linewidth]{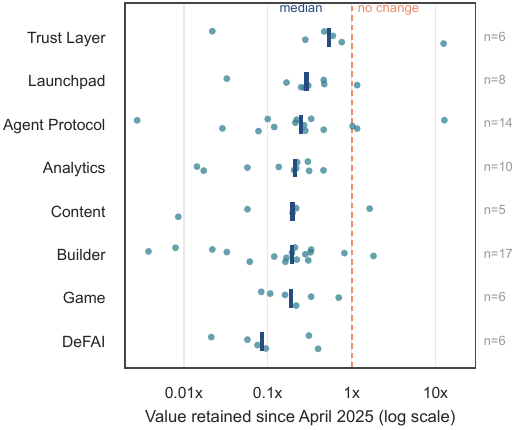}
    \caption{Value retained since \nDateOne by each surviving project, grouped
    by taxonomy subcategory, on a logarithmic scale. The vertical rule marks the
    subcategory median. Retention inside a subcategory spans orders of magnitude,
    which is why the aggregate and the median tell opposite stories.}
    \label{fig:panel_retention}
\end{figure}

Of the \nPanKept still categorized as AI agents, \nPanKeptPriced carry a
capitalization at the cutoff and \nPanKeptUncapitalised are listed without one.
Among the priced survivors, aggregate capitalization fell from \nPanKeptMcapOne
to \nPanKeptMcapTwo, retaining \nPanKeptRetained, but the median project
retained only \nPanKeptMedian. That gap is the finding: \nPanGrew grew,
\nPanLostNinety lost more than 90\% of their value, and \nPanLostNinetyNine lost
more than 99\%. The aggregate is value-weighted, and \nPanHoldingsShare of the
surviving value sits in two holdings: \nPanHoldingOneName, which merely held its
2025 valuation at \nPanHoldingOneRetained, and \nPanHoldingTwoName at
\nPanHoldingTwoRetained. Excluding both, the rest of the set stands at
\nPanExOutlier. Only \nPanTopOneName (\nPanTopOne) and \nPanTopTwoName
(\nPanTopTwo) more than doubled.

Attrition is also uneven across the taxonomy (Figure~\ref{fig:panel_retention}).
The Trust Layer subcategory retained a median \nPanMedianTrust of its value and
Monetization \& Launchpad \nPanMedianLaunchpad, while DeFAI Agents retained
\nPanMedianDeFAI and Content Creation \nPanMedianContent. Measured instead as the
share still capitalized above \nAliveFloor, Trust Layer retains \nPanAliveTrust
and Agent Protocol \nPanAliveProtocol, against \nPanAliveGame for Game \&
Metaverse. The Infrastructure end of the taxonomy proved markedly more durable
than the Creative \& Virtual end. Each subcategory holds at most a few projects,
so these rates order the taxonomy rather than estimate levels;
Figure~\ref{fig:panel_retention} reports the count behind every row.

\observation{At the second cutoff, \nPanKeptShare of the originally capitalized projects are still listed as AI agents, but the median survivor retains only \nPanKeptMedian of its 2025 valuation; aggregate retention of \nPanKeptRetained rests on two holdings carrying \nPanHoldingsShare of the surviving value, and excluding them leaves \nPanExOutlier.}

The case study separates two things a single snapshot cannot. The decline in
aggregate capitalization is not projects failing and leaving; it is projects
persisting at a fraction of their former valuation while new entrants arrive
faster than value does. Survival and success are close to uncorrelated here, and
a study that measured only the survivors' presence would report a healthy
ecosystem.

\begin{answerbox}
\textbf{Answer to RQ1}: The Web3-AI Agent landscape comprises \nPopCount
capitalized tokens worth \nPopMcap, \nClsCount of them classified into four
primary categories. Infrastructure leads in both project count
(\nClsInfraCount) and capitalization (\nClsInfraMcap, \nClsInfraShare of
classified value), while Financial Services (\nClsFinCount) grew fastest between the two
snapshots, from \nPanFinCountShare of assignments to \nClsFinCountShare. Within the archived cohort, aggregate capitalization fell \nCohChange
over \nWindow, and market value is extremely concentrated (Gini \nPopGini,
top decile \nPopTopDecileShare).
\end{answerbox}

%% file: tex/05-rq2.tex
AI agents in Decentralized Finance (DeFi) interact directly with the
programmable layer of blockchains: smart contracts for execution, on-chain
data for verification, and native assets for value transfer. This section
examines the roles these agents play in DeFi operations through current
implementations and case studies from our dataset.

\subsection{Autonomous Trading Strategy Implementation}

A primary role of AI agents in DeFi is the autonomous execution of transactions and complex financial strategies based on user intent or predefined rules. These agents interact directly with blockchain protocols and smart contracts, translating high-level goals into concrete on-chain actions. For instance, platforms like \textit{Griffain}~\cite{griffainGriffain2025} empower users to deploy agents that perform tasks such as setting up recurring token purchases, executing sales when specific market cap targets are met, or automatically depositing idle assets into yield-bearing protocols. Similarly, \textit{Wayfinder}~\cite{wayfinderHomeWayfinder2025} acts as an omnichain protocol where user-owned agents navigate diverse blockchain ecosystems and decentralized applications (DApps). These agents can independently execute transactions ranging from simple token swaps to more intricate strategies like liquidity provision or lending, utilizing dedicated Web3 wallets under user control. This capability moves beyond manual interaction, allowing for persistent, automated management of DeFi activities based on sophisticated logic or real-time triggers, operating with an efficiency often unachievable manually.

\observation{AI agents enable autonomous DeFi strategy execution by translating high-level user intents into complex on-chain actions, moving beyond manual interaction toward automated real-time management with sophisticated logic.}

\subsection{Intelligent Portfolio Construction and Optimization}

AI agents significantly enhance portfolio management within DeFi by offering personalized and adaptive strategies. They analyze a user's transaction history, risk tolerance, and market conditions to construct and dynamically manage DeFi portfolios. \textit{One Click Crypto}~\cite{oneclickHomeBestDeFi2025}, positioning itself as a ``Wealthfront for DeFi'', exemplifies this by using AI to analyze a user's blockchain history and risk profile to recommend and deploy a tailored DeFi portfolio with a single click. Agents can continuously monitor market volatility, protocol risks (e.g., smart contract vulnerabilities, impermanent loss), and yield opportunities. Based on this ongoing analysis, they autonomously rebalance assets, harvest rewards, and shift capital between different protocols or liquidity pools to optimize risk-adjusted returns. 

While not explicitly focused solely on portfolio construction, the agents within ecosystems like \textit{Sharpe AI}~\cite{sharpeaiSharpeAISmart2025} can execute complex, multi-step strategies involved in sophisticated portfolio optimization across different chains and protocols. These AI agents autonomously monitor market conditions, identify optimal entry and exit points, and orchestrate DeFi operations spanning yield farming, liquidity provision and cross-chain arbitrage, guided by user-defined parameters and risk preferences.

These agents serve as intermediaries that translate high-level investment goals into precise technical execution across fragmented DeFi protocols, continuously learning from transaction histories to optimize pathways and strategies while adapting to evolving market dynamics.

\observation{AI agents enhance DeFi portfolio management through personalized analysis of user profiles and continuous optimization, autonomously rebalancing assets and harvesting rewards.}

\subsection{AI-Driven Market Analysis and Intelligence}

AI agents serve as powerful tools for aggregating, processing, and interpreting the vast amount of data generated within the DeFi and broader crypto markets. They provide users with real-time insights, predictive analytics, and sentiment analysis, enabling more informed decision-making. 

Projects like \textit{Aixbt}~\cite{aixbtAIXBT2025} offer a Bloomberg-style intelligence dashboard driven by AI, delivering real-time research on tokenomics, narrative tracking, and smart money flows faster than human analysts. \textit{Assemble AI}'s agent, NS3~\cite{assembleaiAssembleAIAI2025}, focuses on Web3 journalism, using advanced AI reasoning to analyze crypto and economic trends, market psychology, and potential ripple effects, delivering insights in multiple languages. \textit{DexCheck AI}~\cite{dexcheckaiDexCheckAIAI2025} leverages AI and machine learning for real-time DEX and NFT market analysis, featuring an AI-powered on-chain search engine and integration with Telegram trading bots for automated solutions. Furthermore, \textit{Hey Anon}~\cite{heyanonHeyAnon2025} combines conversational AI with real-time data aggregation, allowing users to query project updates and analyze trends across multiple platforms. These intelligence agents sift through noise, identify patterns, and deliver actionable information, making the complex and fast-moving DeFi market more understandable.

\observation{AI agents process vast DeFi market data to deliver real-time insights and predictive analytics that enable faster, more informed decision-making than human analysis.}

\subsection{Improving DeFi Accessibility and Interaction}

A crucial role for AI agents is simplifying the complex user experience associated with DeFi and Web3 interactions. By reducing technical complexity, agents make DeFi more accessible to a broader audience. \textit{Hey Anon}~\cite{heyanonHeyAnon2025}, for example, uses conversational AI to allow users to manage DeFi operations through natural language prompts, simplifying tasks and information retrieval. \textit{Wayfinder} explicitly aims to improve blockchain accessibility via its natural language interface, enabling both novice and experienced users to execute complex cross-chain transactions (like bridging assets) through generalized instructions, which the AI agent translates into the necessary multi-step on-chain actions. These agents act as intelligent interfaces, managing wallet interactions, gas fees, and protocol specifics behind the scenes. Wayfinder's concept of agents learning from past interactions and leveraging shared knowledge ("Wayfinding Paths," network memory) further points towards systems that adapt to user needs and continuously improve the interaction experience within decentralized ecosystems~\cite{wayfinderHomeWayfinder2025}. This focus on usability is critical for driving wider adoption of DeFi technologies.

\observation{AI agents improve DeFi accessibility through natural language interfaces that simplify complex operations and reduce technical barriers for mainstream users.}

\subsection{Agent-Native Payments: The x402 Settlement Layer}
\label{subsec:x402}

The DeFi integrations surveyed above all assume an agent acting on behalf of a
human account. A different pattern has emerged in 2025--2026:
protocols in which the agent is itself the paying party, settling
machine-to-machine obligations without a human in the transaction path. The
x402 protocol, which revives the dormant HTTP 402 ``Payment Required'' status
code and settles in stablecoins, is the most widely adopted of
these.\footnote{Introduced by Coinbase in May 2025; governance moved to the x402
Foundation under the Linux Foundation on 2 April 2026.}

\begin{figure}[t]
    \centering
    \includegraphics[width=\linewidth]{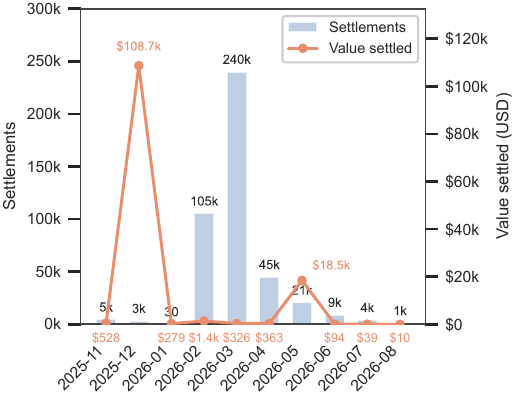}
    \caption{x402 \texttt{SettlementRouter} activity on Base by month. Volume
    and value move independently because the two regimes are different kinds of
    payment: \nXFourDecMonth's \nXFourDecValue arrives in
    \nXFourDecSettlements human-scale transfers (median \nXFourDecMedian,
    largest \nXFourDecLargest), while \nXFourPeakMonth's \nXFourPeak transfers
    are machine-scale (median \nXFourPeakMedian, largest \nXFourPeakLargest)
    and total \nXFourPeakValue.}
    \label{fig:x402}
\end{figure}

To measure adoption rather than announcement, we queried the on-chain
settlement record directly. Through our \nDateTwo cutoff, the x402
\nXFourContract on \nXFourChain recorded \nXFourSettlements settlements between
\nXFourPayers distinct payers and \nXFourPayees distinct payees, all denominated
in \nXFourAsset.\footnote{This is one x402 implementation on one chain, and
bounds what follows in two ways. Within \nXFourChain, traffic settled by
facilitators through direct USDC \texttt{transferWithAuthorization} bypasses
the \nXFourContract and is not captured. Across chains, \nXFourChain is the
only chain for which decoded x402 settlement data exists, so deployments
elsewhere are outside the measurement rather than shown to be absent. Both make
these figures a lower bound, and every x402 result below describes the
\nXFourChain settlement record specifically.} Activity begins in
\nXFourFirstMonth, rises to a peak of \nXFourPeak settlements in
\nXFourPeakMonth, and then falls monotonically to \nXFourLastSettlements in
\nXFourLastMonth, \nXFourDeclineShare of the peak (Figure~\ref{fig:x402}).
Adoption of the standard has not yet produced a durable payment stream.

The economic magnitude of that activity is a different matter. The
\nXFourSettlements settlements move \nXFourValue in total. The median payment is
\nXFourMedian, the 90th percentile \nXFourPNinety, and
\nXFourUnderCentShare of all settlements transfer less than one cent;
\nXFourUnderDollarShare transfer less than one dollar. The arithmetic mean of
\nXFourMean is not a meaningful summary: a handful of dollar-scale transfers,
the largest \nXFourLargest, sets it in a record of sub-cent ones. The supply
side is also markedly more concentrated than the demand side: from
\nXFourSustainedFrom onward, monthly payers run from \nXFourPayersLow to
\nXFourPayersHigh while the payees in any month never exceed
\nXFourPayeesHigh.

The apparent contrast in Figure~\ref{fig:x402}, \nXFourDecMonth's
\nXFourDecValue against \nXFourPeakMonth's \nXFourPeakValue, is not a contrast
between real and negligible economic activity. Nearly all of that value arrived
on a single day (\nXFourBurstDate, \nXFourBurstSettlements settlements totalling
\nXFourBurstValue), and the on-chain structure of that day is a cascading split:
a \nXFourBurstSeed seed is halved at each hop
(\$9{,}997 $\to$ \$4{,}999 $\to$ \$2{,}499 $\to$ \$1{,}250), with each
recipient forwarding at least 95\% of what it received.\footnote{%
Of the \nXFourBurstAddresses addresses active that day, \nXFourBurstRelays both
received and sent. Among those, addresses that forwarded $\geq$\,95\% of their
inflow account for at least \nXFourBurstForwarded
(\nXFourBurstForwardedShare of the day's volume), a conservative lower bound
because a separate cluster of \nXFourBurstUniform uniform
\nXFourBurstUniformValue payments from distinct payers is not captured by this
criterion.} At least \nXFourBurstForwardedShare of that day's
reported volume is the same dollars counted repeatedly as they traverse the
tree, and the true figure is likely higher. The correct reading is therefore
not that December had genuine economic activity and March did not, but that
neither month did: December's dollar volume is self-cycling, and March's is
sub-cent machine calls. One further month reaches dollar scale,
\nXFourSecondMonth, at \nXFourSecondValue over \nXFourSecondSettlements
settlements. Outside these the \nXFourChain record moves less than a cent per
settlement.

\observation{Agent-native payment infrastructure is operational at genuine transaction scale (\nXFourSettlements settlements on \nXFourChain) but economically negligible: most of the \nXFourValue those settlements move arrives in a single day of self-cycling transfers, and the sustained regime that follows moves sub-cent amounts per settlement.}

This parallels the identity-layer finding of Section~\ref{subsec:sybil}, where
permissionless registration inflates registry counts through two structurally
distinct patterns. Here, permissionless settlement inflates volume through
cascading forwarding. In both cases the metric that appears healthiest
(registrations, transaction count, dollar volume) is the one most susceptible
to mechanical amplification, and the metric that resists amplification
(distinct economic counterparties, median transfer value) tells the opposite
story. Combined with the token-layer result of
Section~\ref{subsec:longitudinal}, where a multiplying token count accompanied
a decline in aggregate capitalization, the pattern replicates
independently at three layers: \textit{activity without capitalization}. The
Web3-AI agent ecosystem's infrastructure adoption is substantially ahead of
its economic adoption, and measurements taken at any single layer would be
liable to mistake one for the other.

Across these roles, AI agents extend DeFi from manual, single-transaction
interaction to persistent, multi-step automation, and the x402 settlement
record shows that the infrastructure for agent-native payments exists even
though its economic throughput remains negligible.

\begin{answerbox}
\textbf{Answer to RQ2}: AI agents transform DeFi through four key roles: autonomous trading strategy implementation, intelligent portfolio construction and optimization, AI-driven market analysis and intelligence, and improving accessibility through natural language interfaces. These agents enable persistent automation, personalized strategies, real-time insights, and simplified interactions, enhancing DeFi efficiency and accessibility.
\end{answerbox}

%% file: tex/06-rq3.tex
AI agents enhance Web3 governance by addressing critical challenges throughout the governance lifecycle~\cite{chafferDecentralizedGovernanceAutonomous2025}. Building upon Ostrom's established governance theory~\cite{ostrom1990governing}, we analyze AI agent integration across three fundamental governance phases: \textit{decision-making processes} (proposal analysis and community engagement), \textit{implementation and enforcement} (automated monitoring and execution), and \textit{system evolution} (adaptive mechanism design~\cite{harreFirmsComputationAI2025}).

\subsection{Proposal Analysis and Community Engagement}

AI agents streamline governance processes by automating proposal analysis and enhancing community engagement capabilities. These agents automatically parse complex proposals, provide key summaries, and identify implications across economic, security, and community alignment dimensions. For example, agents developed on platforms like \textit{Fetch.ai}~\cite{aiFetchaiBuildDiscover2025} or utilizing analytical capabilities similar to \textit{ChainGPT}~\cite{chaingptExploreAIAgents2025} analyze smart contract modifications within proposals and flag potential security vulnerabilities before voting commences.

AI agents enhance deliberation quality by facilitating sophisticated voting strategies and community engagement. Users delegate voting power to AI agents programmed with specific preferences or ethical frameworks, leveraging ecosystems like \textit{Autonolas (Olas)}~\cite{olasOlasCoownAI2025} for co-owned agent services or \textit{Artificial Liquid Intelligence (ALI)}~\cite{artificialliquidintelligenceIntroductionAINews2025} where agents execute decisions based on token-holder defined rules.

Community accessibility improves through AI agents that translate complex technical terminologies into simplified language and provide personalized governance notifications. Agents built using frameworks from \textit{MyShell}~\cite{myshellMyShellAIBuild2025} create interfaces that lower barriers for less technical community members. These agents monitor discussion forums, summarize key arguments, track community sentiment, and identify emerging consensus points.

How much of this is deployed rather than designed? No governance category
emerged from the taxonomy, so we coded the population directly. Of the
\nClsCount classified projects, \nGovDescribed carry a usable description;
\nGovVocabulary of those (\nGovVocabularyShare) mention governance, a DAO,
voting or proposals at all, and \nGovWithAgent do so in a sentence that also
names an agent. Reading those \nGovWithAgent separates three things a keyword
count conflates. \nGovUtilityCount are the standard token-utility
construction, in which the token confers voting rights on its holders and the
agent is what the protocol sells, not what governs it. \nGovSubjectCount
describe governance \textit{of} agents (rules constraining what an agent is
permitted to do, as in Quack AI's governance layer or the Covenant
Foundation's agent governance framework) rather than governance \textit{by}
them. \nGovActorCount describe agents acting as governance participants:
Autonolas' \textit{Governatooorr} delegate, ai16z's agent-led venture DAO,
Thirdfy's agent ``CEO'', SentismAI's autonomous DAO operation, Chirper's
agentic DAOs, JuliaOS' cross-chain governance agents, and Owocki's debating
``AI politicians''. Those \nGovActorCount are together worth \nGovActorMcap,
\nGovActorShare of classified capitalization, but \nGovActorLargest, a
general-purpose L1 whose description gestures at agentic organizations,
accounts for almost all of it; excluding it, agents that govern are backed by
\nGovActorMcapExLargest, well under one percent of the population's value.

\observation{\nGovActorCount of \nClsCount classified projects describe agents
acting as governance participants, and excluding a single general-purpose L1
they hold \nGovActorMcapExLargest of \nClsMcap in classified capitalization.
Governance vocabulary is common in this population but almost always denotes
voting rights attached to a token rather than an agent that votes.}

\observation{Platforms in this population offer agent-based proposal analysis, security monitoring and natural-language governance interfaces. These are advertised product capabilities; adoption is measured separately below.}

\subsection{Automated Monitoring and Enforcement of Governance Decisions}

AI agents ensure faithful execution of approved governance proposals through continuous monitoring of on-chain activities and relevant off-chain data sources. These agents verify that decisions encoded in passed proposals are correctly enacted by tracking smart contract interactions and balance changes. For example, when a proposal mandates fee structure modifications or treasury disbursements, AI agents could leverage \textit{OriginTrail}'s Decentralized Knowledge Graph to monitor relevant smart contract executions to confirm compliance.

Automated oversight mechanisms flag deviations from approved proposals and trigger appropriate responses. Monitoring agents automatically alert communities or designated oversight bodies when violations are detected, and can even trigger pre-defined contingency actions via smart contracts. This capability enhances accountability and transparency in post-voting governance phases. The reliability of such agents depends on robust \textit{Agent Protocol \& DePIN} infrastructure, such as that developed by \textit{Autonolas (Olas)}, and \textit{Trust Layer} mechanisms to ensure operational integrity and data veracity. AI agents within the \textit{Humans.ai} ecosystem exemplify this approach by emphasizing ethical AI governance with transparent and accountable operations.

\observation{Agent-based monitoring is the mechanism proposed for post-vote compliance: continuous observation of on-chain execution, verification against the passed proposal, and alerting on deviation. We found no project in the classified population reporting it in production.}

\subsection{Adaptive Governance and Mechanism Design}

AI agents contribute to dynamic and resilient governance models by analyzing historical governance data, participation patterns, and network performance metrics. These agents identify inefficiencies in existing governance mechanisms and simulate potential impacts of proposed changes to voting rules, quorum thresholds, or incentive structures. For instance, AI tools specialized from \textit{ChainGPT}'s code analysis capabilities analyze governance patterns and model outcome scenarios for proposed mechanism modifications~\cite{chaingptExploreAIAgents2025}.

Iterative governance refinement occurs through AI-assisted analysis and recommendation systems. AI agents suggest adjustments to delegation incentives for improved voter participation or propose modifications to proposal lifecycles that enhance decision-making speed without compromising deliberation quality. 

Advanced AI models supported by \textit{Infrastructure} projects like \textit{NEAR Protocol} and \textit{Fetch.ai}'s Open Economic Framework enable sophisticated governance mechanism design. These systems become more responsive and better aligned with long-term community objectives through continuous learning and adaptation. AI agents from \textit{Virtuals Protocol}~\cite{virtualVirtualsProtocolSociety2025}, designed for human-curated AI, incorporate community feedback into learning processes to develop adaptive governance models that evolve with community needs.

\observation{Adaptive governance, in which agents analyse participation history to recommend changes to voting rules or incentives, appears in vendor material for several infrastructure projects and in no deployment we could verify.}

The tooling for agent-mediated governance exists across proposal analysis,
monitoring and mechanism design, but the gap between advertised capability and
measured adoption is wide. Closing it will require addressing AI
decision-making transparency, algorithmic bias and agent security, challenges
that the current population has largely deferred.

\begin{answerbox}
\textbf{Answer to RQ3}: AI agents address Web3 governance along three axes
(proposal analysis and community engagement, automated monitoring and
enforcement, and adaptive mechanism design) and platforms exist for each.
Deployment is the exception rather than the rule: \nGovActorCount of
\nClsCount classified projects describe agents acting as governance
participants, and net of one general-purpose L1 they hold
\nGovActorMcapExLargest of \nClsMcap in classified capitalization. RQ3
therefore reports a designed
capability with a measured adoption floor, not an established practice.
\end{answerbox}

%% file: tex/07-rq4.tex
AI agents extend Web3 security from static, rule-based analysis to adaptive
vulnerability detection. By combining natural language processing, machine
learning and automated reasoning, they identify complex security
vulnerabilities that conventional tools often miss~\cite{wang2023gptlens}.
This section compares traditional approaches with agent-enhanced solutions.

\subsection{Traditional Web3 Security Approaches}

Before the emergence of LLM-driven agents, automated security auditing in Web3 ecosystems primarily relied on rule-based analysis methods~\cite{nikolicFindingGreedyProdigal2018,tsankovSecurifyPracticalSecurity2018,feistSlitherStaticAnalysis2019}. These approaches, while foundational, exhibit significant limitations that prevent them from replacing human auditors~\cite{chenWhenChatGPTMeets2024}.

\paratitle{Rule-Based Security Techniques}Static analysis tools, such as Slither~\cite{feistSlitherStaticAnalysis2019}, examine code without execution to detect common vulnerabilities, including reentrancy attacks and integer overflows, through predefined pattern matching. However, they suffer from high false positive rates and struggle to comprehend complex business logic~\cite{chenFORGELLMdrivenFramework2025}. Dynamic analysis frameworks, such as Mythril~\cite{consensysConsensysMythrilMythril2018} and Manticore~\cite{mossbergManticoreUserFriendlySymbolic2019}, utilize symbolic execution to explore runtime behavior and identify security flaws; however, they face scalability challenges and state explosion problems with complex contracts~\cite{chaliasosSmartContractDeFi2024}. Formal verification approaches, such as Certora~\cite{CertoraProverDocumentation} and KEVM~\cite{runtimeverificationKEVMEvmsemantics2025}, mathematically prove contract correctness through theorem proving and model checking, providing rigorous security guarantees but requiring extensive manual specification writing and struggling to adapt to evolving vulnerability patterns~\cite{chenWhenChatGPTMeets2024}. 

\paratitle{Data-Driven Analysis Techniques}Previous Web3 vulnerability detection efforts extensively utilized machine learning (ML) and deep learning (DL) approaches~\cite{wangEthereumSmartContract2021}. These methods typically train models on pre-constructed, manually annotated Web3 vulnerability datasets to detect known vulnerability patterns. Tools like VulDeePecker~\cite{zouVulDeePeckerDeepLearningBased2021} and DR-GCN~\cite{zhuangSmartContractVulnerability2020} apply neural networks to identify code patterns associated with security flaws. However, these approaches consistently face generalization limitations and maintenance challenges, struggling to detect vulnerabilities outside their training distributions and failing to adapt to emerging Web3 attack vectors without significant retraining efforts~\cite{liVulnerabilityDetectionFinegrained2021,shenIntelliConConfidenceBasedApproach2024}.

\observation{Traditional Web3 security approaches face limitations in generalization, adaptability to emerging threats, and comprehensive business logic analysis.}

\subsection{AI Agent-Enabled Web3 Security Technologies}

AI agents enhance smart contract auditing through intelligent analysis that identifies sophisticated vulnerabilities and adapts to evolving attack patterns. LLMs analyze smart contract code with human-like reasoning, extending beyond surface-level vulnerabilities to identify complex logical issues through business logic comprehension~\cite{zhengUnderstandingLargeLanguage2024a}.

\paratitle{Agent-Based Detection Systems}\textit{GPTScan}~\cite{sun2024detecting} and its commercial implementation, MetaScan~\cite{metatrustMetaScanAutomatedAIPowered2025}, address the critical gap where approximately 80\% of Web3 security bugs cannot be audited by existing tools through a three-phase methodology: vulnerability decomposition, GPT-based matching, and static confirmation. The system achieves over 90\% precision for token contracts while detecting nine previously unknown vulnerabilities missed by human auditors, completing analysis in 14.39 seconds per 1,000 lines of Solidity code at \$0.01 USD per scan.

Advanced AI security systems employ multiple specialized agents working collaboratively to enhance vulnerability detection accuracy. For example, \textit{iAudit}~\cite{ma2025Combining} demonstrates this approach through its two-stage fine-tuning framework using specialized LLM agents: a Detector model for initial vulnerability identification and a Reasoner model for generating explanations, with additional Ranker and Critic agents iteratively selecting vulnerability causes and debating explanations, achieving 91.21\% F1 score and 91.11\% accuracy on real smart contract vulnerabilities. This collaborative approach mimics human expert auditing while maintaining consistency and eliminating fatigue-related errors.

\paratitle{AI-Powered Security Intelligence Platforms}
AI agents extend beyond individual vulnerability detection to create comprehensive security intelligence platforms providing real-time threat monitoring across Web3 ecosystems. ChainGPT's integrated platform demonstrates this evolution through multiple AI-powered tools: Smart Contract Auditor delivers automated 6-step auditing within 1-2 minutes, CryptoGuard browser extension provides real-time phishing detection, and ChainAware.ai reports 98\% accuracy in fraud detection through continuous transaction monitoring~\cite{chainawareaiChainAwareaiWeb3AI2025}. These platforms integrate with multiple threat intelligence providers, including Forta~\cite{fortaForta2025} and GoPlus~\cite{goplusOverviewGoPlusSecurity2025}, enabling comprehensive cross-platform security analysis while maintaining real-time monitoring capabilities across various blockchain architectures.

\observation{Agent-driven Web3 security overcomes traditional limitations by understanding real business logic, collaborative multi-agent systems that achieve high accuracy, and monitoring platforms that adapt to evolving threats.}

\begin{answerbox}
\textbf{Answer to RQ4}: AI agents enhance Web3 security through automated
smart contract auditing and security intelligence platforms, addressing
traditional auditing limitations by identifying complex logic vulnerabilities
at a cost per scan that makes continuous coverage practical.
\end{answerbox}

%% file: tex/08-rq5.tex
Web3 supplies trust for AI agents through verifiable execution, on-chain identity, and public accountability. Individual projects first built these mechanisms separately. ERC-8004 now standardizes them as three registries: a Validation registry for verifying agent work, an Identity registry for naming the agent, and a Reputation registry for feedback about it. This section follows those three layers and measures the two that carry on-chain activity.

\subsection{Verifiable Execution}

Trusted Execution Environments and advanced cryptographic protocols form the foundation for secure AI agent operations~\cite{salehBlockchainSecureDecentralized2024}. Projects like \textit{Mind Network} utilize Fully Homomorphic Encryption to enable AI agents to perform computations on encrypted data while preserving privacy. The integration of \textit{ai16z (Eliza)} framework~\cite{waltersElizaWeb3Friendly2025a} with \textit{Phala Network}'s TEE infrastructure~\cite{phalaPhalaNetwork2025} makes agent actions verifiable at execution time. Multi-party cryptography and game theory mechanisms in \textit{Fetch.ai}'s architecture~\cite{aiFetchaiBuildDiscover2025} enable secure coordination between agents without requiring mutual trust. Consensus mechanisms such as \textit{NEAR Protocol}'s \textit{Doomslug}~\cite{nearBlockchainAI2025} provide distributed verification for AI operations, while \textit{OriginTrail}'s Decentralized Knowledge Graph~\cite{origintrailOriginTrailPoweringShift2025} enables agents to access and contribute to shared, verified knowledge bases.

ERC-8004's Validation registry standardizes posting verifications of agent work on-chain, so verification outcomes are composable and queryable by other contracts, replacing per-project implementations with a shared addressing layer.

\observation{TEE, FHE and consensus-based verification let AI agents compute on sensitive data with verifiable execution, and ERC-8004's Validation registry turns these per-project mechanisms into a shared on-chain record.}

\subsection{Standardized Agent Identity as a Trust Primitive}
\label{subsec:sybil}

The mechanisms above attach trust to an agent's execution. None answers which agent is executing. Before ERC-8004, each project filled this gap with a bespoke mechanism. ERC-8004 fills it with a shared standard. Deployed to mainnet on 29 January 2026, it is the first widely
adopted on-chain standard for agent identity
(Section~\ref{subsec:standards}).

What it delivers is composability. An ERC-8004 Identity registration is an ERC-721 token with a resolvable metadata URI, so an agent's identity is transferable, ownable, queryable by any contract, and portable across chains. The Reputation and Validation registries extend the same addressing to third-party attestations about the agent and to verifications of its work, so a consuming agent can in principle resolve an unknown counterparty to a record without trusting the counterparty's own claims. Standardizing identity also standardizes the target: a registry entry is only as meaningful as the cost of creating one, and ERC-8004 registration is permissionless.

We measured the resulting registry population across every chain with decoded registry data available, Ethereum mainnet included.\footnote{Mainnet is decoded under a separate namespace from the other fifteen chains and has to be queried separately, but it is the same contract: the registry is deployed at the same address on all sixteen.} Through \nDateTwo, the Identity registry recorded \nErcRegistrations registrations from \nErcRegistrants distinct registrant addresses over \nErcChains chains.\footnote{Registrants are counted per chain, so the cross-chain total is an upper bound on unique entities.} \nErcTopChain accounts for \nErcBNBRegistrations registrations, Base for \nErcBaseRegistrations and Ethereum mainnet for \nErcEthereumRegistrations, with the remaining \nErcTailChains chains contributing \nErcTail between them. Adoption is compressed rather than gradual: Ethereum, Base and Monad all begin on the mainnet launch date itself, and every other chain follows within two weeks.

\begin{figure}[t]
    \centering
    \includegraphics[width=\linewidth]{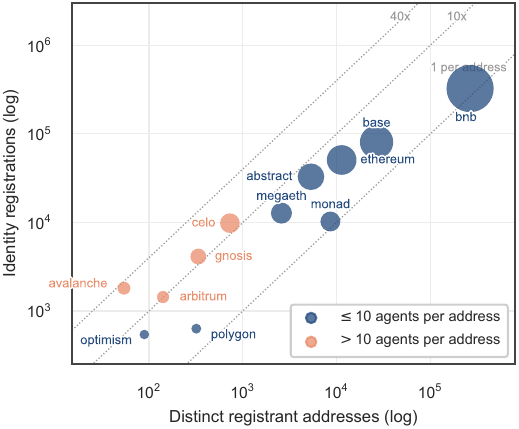}
    \caption{ERC-8004 Identity registrations by chain, against registrations per
    distinct registrant. The two series move inversely: the highest-volume chains
    have the lowest concentration, and the most concentrated chains are small.}
    \label{fig:erc8004}
\end{figure}

The ratio of registrations to distinct registrants varies by more than an order of magnitude across chains (Figure~\ref{fig:erc8004}). On Monad the ratio is \nErcMonadRatio (\nErcMonadRegistrants addresses registering almost exactly one agent each), and on BNB Chain \nErcBNBRatio across \nErcBNBRegistrants addresses. On Celo the ratio is \nErcCeloRatio from only \nErcCeloRegistrants addresses, and on Avalanche \nErcAvalancheRatio from \nErcAvalancheRegistrants. Ethereum mainnet, by an order of magnitude the most expensive chain to register on, sits between the two regimes at \nErcEthereumRatio and is still third by volume: registration cost bounds the count without distinguishing the patterns. These are not the same phenomenon at different intensities. The first is a \textit{many wallets, one agent} pattern consistent with airdrop farming, where the addresses are the disposable resource; the second is a \textit{one wallet, many agents} pattern where the registry entries are. Both inflate registry counts, but they defeat different countermeasures. Per-address rate limiting is ineffective against the first, while proof-of-personhood or registration fees are ineffective against the second.

\observation{Permissionless agent identity registries exhibit two structurally distinct inflation patterns (many single-use addresses each registering one agent, and a few addresses each registering dozens) which defeat different countermeasures and should not be measured as one phenomenon.}

\subsection{Accountability and Reputation}

Web3's transparent governance enables community oversight of AI agent behavior through blockchain-based systems. Immutable on-chain audit trails record agent actions for later accountability analysis. \textit{Commune AI}'s protocol~\cite{communeaiCommuneAI2025} demonstrates community-driven validation through permissionless frameworks that enable stakeholder oversight while maintaining operational efficiency.

Moving agent identity on-chain relocates the trust problem rather than resolving it. A registry entry provides a durable record that an identity was created; it provides no evidence that the identity corresponds to a distinct operator, or that the agent behind it does anything at all. That evidence would come from the Reputation registry ERC-8004 defines alongside identity. Through \nDateTwo, the Reputation registry recorded \nErcRepCalls \texttt{giveFeedback} calls, a volume comparable to the \nErcRegistrations identity registrations and, on Base, exceeding registrations by a factor of \nErcRepBaseMultiple. The difficulty is in who produces it. Those feedback entries come from only \nErcRepSenders distinct addresses, \nErcRepSenderShare as many as the addresses that registered agents. On BNB Chain, \nErcRepBNBSenders addresses account for \nErcRepBNBCalls entries; the entire reputation state of the Gnosis registry was written by a single address.

A reputation registry whose supply of feedback is controlled by a few dozen addresses offers no more assurance than an empty one. The identity layer's Sybil exposure is visible in registration counts, but the reputation layer's is visible only in the concentration of its writers, a property that a client checking whether an agent has feedback would not observe.

\observation{Standardized on-chain identity solves agent addressability, not agent trustworthiness: because registration and attestation are both permissionless and inexpensive, ERC-8004's registries certify that a record exists rather than that it means anything.}

The productive direction is not further identity standardization but making identity and attestation costly in ways that are verifiable on-chain: staking or bonding against an agent's registration, proof-of-personhood gating on attestation, or validation attested by the trusted execution environments discussed above rather than by an unconstrained address. Each of these composes with ERC-8004 rather than replacing it; the addressing layer is settled, and the open problem has narrowed to the economics layered on top of it.

\begin{answerbox}
\textbf{Answer to RQ5}: Web3 gives AI agents verifiable execution, and ERC-8004 standardizes it together with identity and reputation. Because registration and attestation are permissionless, the registries certify that records exist, not that they mean anything. The open problem is making identities and attestations costly enough to be informative.
\end{answerbox}

%% file: tex/09-discussion.tex
The landscape analysis and on-chain measurements point to a consistent
pattern: the Web3-AI agent ecosystem's infrastructure is substantially ahead
of its economic adoption, and the standards layer that emerged in 2025--2026
has shifted several open problems from protocol design to incentive design.
This section discusses the research directions those shifts open and the
technical challenges that remain.

\subsection{Emerging Research Directions}

\paratitle{Agent Memory and Context Persistence}
Current AI agents face limitations in maintaining long-term memory and contextual understanding across extended interactions~\cite{liuLostMiddleHow2024,zhangziyaoLLMHallucinationsPractical2025}. Web3 infrastructure offers novel solutions through decentralized storage and blockchain-based state persistence. Decentralized storage networks, such as IPFS~\cite{benetIPFSContentAddressed2014}, provide tamper-proof memory systems that enable agents to maintain coherent, long-term contexts while ensuring data integrity and availability. Research opportunities include developing Web3-native memory architectures that enable agents to maintain verifiable interaction histories, learn from past experiences, and build persistent knowledge graphs. 

\paratitle{Agent Decentralized Identity (Agent DID)}
Sovereign agent identity has been an open problem: agents operated through
user-delegated permissions, limiting their autonomy and creating a dependency
on human wallet infrastructure~\cite{nagwareVerifiableAIHow2025,9031542}.
ERC-8004 has supplied the missing primitive, and our measurements
(Section~\ref{subsec:sybil}) show it is being used at scale. The open problem is
therefore no longer the absence of an identity standard but the absence of cost
in it. Permissionless registration produces two distinct inflation patterns, and
permissionless attestation produces a reputation layer whose supply is
controlled by a small number of addresses. The research priority is now
mechanisms that make identity and attestation verifiably expensive (bonded
registration, proof-of-personhood gating, validation attested by trusted
execution environments), together with certification and liability frameworks
for agent-controlled operations, which remain genuinely unaddressed.

\paratitle{Portable AI Agent Digital Assets}
Agent portability is now partially standardized. An ERC-8004 identity is an
ERC-721 token, so an agent's identity and its accumulated attestations are
transferable and ownable by default, and account abstraction supplies the
wallet architecture: ERC-4337 for programmable agent accounts and EIP-7702 for
scoped, revocable authority delegated from an existing account. The gap has
moved from specification to adoption. Account abstraction is almost entirely
absent from the agent codebases we surveyed (\nStdAaRepos repositories in a
pool of \nStdPool reference ERC-4337 or EIP-7702 at all), so agents overwhelmingly still
transact from bare keys or from their operator's wallet. Why a standardized,
widely available account model is not being adopted by the projects that most
need it is the more tractable research question.

\paratitle{Decentralized Multi-Agent Coordination}
Agent-to-agent communication has likewise standardized on
A2A~\cite{googleAnnouncingAgent2AgentProtocol2025} and MCP for tool access, with
ERC-8004 supplying the trust layer A2A leaves unspecified and x402 and AP2 the
settlement and authorization layers. Coordination protocols are consequently no
longer the bottleneck. What our measurements suggest is the bottleneck instead
is economic: on the one chain where x402 settlement can be observed, the
machine-payment layer settles hundreds of thousands of transactions against a
few dozen service endpoints, moving negligible value
(Section~\ref{subsec:x402}). Decentralized coordination among agents is
technically operational and commercially inert. The open research questions are
accordingly about incentive design (what makes an agent service worth paying
for, and how a coordinating agent selects among providers whose reputation
signals are cheap to manufacture) rather than about protocol design.
Regulatory frameworks for cross-border agent coordination remain open.

\paratitle{Real World Assets (RWA) Integration with AI Agents}
The convergence of RWA tokenization and AI agents represents an underexplored frontier in Web3 finance. Despite the RWA market reaching \$23 billion in the first half of 2025 with 260\% growth~\cite{vardaiRWATokenMarket2025}, AI agent integration remains limited. Current implementations face challenges including centralization risks in off-chain oracle systems~\cite{chenExploringSecurityIssues2024}, low efficiency in asset verification processes, and interoperability issues between RWA protocols and AI agents. These limitations significantly increase development costs and complexity. Research opportunities include developing autonomous asset management systems, creating AI-driven risk assessment frameworks for tokenized assets, and establishing protocols for agent-controlled real-world asset operations.

\subsection{Technical Challenges and Limitations}

Current AI agent implementations in Web3 environments face several fundamental limitations that constrain their effectiveness and adoption.

\paratitle{AI Agent Reliability Issues}
Contemporary AI agents exhibit significant reliability challenges that limit their suitability for autonomous Web3 operations. Hallucination remains a critical problem, where agents generate false information or make incorrect decisions based on fabricated data~\cite{zhangziyaoLLMHallucinationsPractical2025}. Additionally, limited context memory restricts agents' ability to maintain coherent, long-term interactions, particularly in complex financial or governance operations that require extended reasoning chains~\cite{zhengUnderstandingLargeLanguage2024a}. High computational costs make real-time agent operations expensive and potentially unsustainable for resource-constrained applications~\cite{zhaoSurveyLargeLanguage2025}.

\paratitle{Security Vulnerabilities}
AI agents face increasing security threats that pose substantial risks for Web3 integration. Prompt injection attacks enable malicious actors to manipulate agent behavior by crafting inputs that override intended instructions~\cite{songProtocolUnveilingAttack2025}. Jailbreaking techniques enable attackers to bypass security measures and cause agents to perform unauthorized actions~\cite{chenRMCBenchBenchmarkingLarge2024}. These vulnerabilities are particularly concerning in Web3 contexts where agents may control significant financial assets or participate in irreversible blockchain transactions.

\paratitle{Trust and Adoption Barriers}
User trust in AI agents remains a significant barrier to adoption, particularly in high-stakes financial applications. Recent research~\cite{songProtocolUnveilingAttack2025} indicates that users express significant reluctance to delegate cryptocurrency asset management to AI agents due to concerns about reliability, security, and accountability. This trust deficit creates a fundamental barrier to the widespread adoption of AI agents, as users prefer to maintain direct control over financial decisions rather than rely on autonomous systems.

%% file: tex/10-related_work.tex
\subsection{Blockchain and AI Integration}

The convergence of blockchain and artificial intelligence technologies has become a significant research focus across multiple domains. Bhumichai \etal~\cite{goyal2024convergence} provide a comprehensive review of AI-blockchain convergence, identifying key research directions and challenges in combining these technologies. Choi \etal~\cite{raza2024technological} present a comprehensive survey categorizing research into two main scenarios: using blockchain to enhance AI capabilities and applying AI to advance blockchain technology. Kayikci \etal~\cite{kayikciBlockchainMeetsMachine2024} focus specifically on machine learning and blockchain integration, examining how blockchain provides secure and transparent transaction recording while machine learning enables data-driven decision-making through large-scale data analysis.

Our work differs from these general surveys in that it specifically examines AI agents as autonomous entities operating within Web3 ecosystems, rather than treating AI as a complementary technology to blockchain.

\subsection{AI Agents in Decentralized Systems}

Research on AI agents operating within decentralized systems has explored multi-agent coordination, autonomous economic systems~\cite{nartey2024decentralized}, and Decentralized Autonomous Organizations (DAOs). Karim \etal~\cite{karimAIAgentsMeet2025} survey secure and scalable collaboration among multi-agents in blockchain environments, focusing on GenAI and LLM-based agents that represent the forefront of intelligent systems in decentralized environments. Ante~\cite{ante2024autonomous} investigates autonomous AI agents specifically within decentralized finance, developing a typology based on a qualitative analysis of 306 major crypto AI agents and introducing a quadrant-based framework that distinguishes four archetypal system configurations. Chaffer \etal~\cite{chafferDecentralizedGovernanceAutonomous2025} examine decentralized governance frameworks for autonomous AI agents, while Ballandies \etal~\cite{ballandiesDAOsCollectiveIntelligence2024} investigate collective intelligence mechanisms in decentralized autonomous organizations.

While previous work primarily focuses on theoretical frameworks or limited application scenarios, our research extends by providing a systematic market analysis of real-world Web3-AI Agent implementations across diverse application domains, revealing actual deployment patterns and market dynamics that existing literature has not captured.

\paratitle{Measurement of the agent standards layer} A parallel line of work has
begun measuring the standards layer directly. Most closely related, an empirical
study of the ERC-8004 ecosystem~\cite{canTrustlessAgentsBeTrusted2026} analyzes
roughly 173{,}000 agents registered on Ethereum, BSC and Base between January
and May 2026, and a companion dataset paper~\cite{earlyBlockchainRegisteredAgents2026}
publishes the early Ethereum registrations. Our ERC-8004 measurements
(Section~\ref{subsec:sybil}) cover the three chains
they study and extend the census to all sixteen with decoded registry data,
but the contribution we claim is different in kind. Those studies take a single standard as their unit of
analysis; ours takes the ecosystem, and the standards layer enters as one
dimension among the five research questions, measured alongside the token
population that preceded it and the payment layer that grew beside it.

We accordingly position this work as the broadest available landscape of the Web3-AI agent ecosystem: \nPopCount tokens and two independent sampling frames, five analytical dimensions and two timepoints \nWindow apart.

%% file: tex/11-conclusion.tex
This paper presents the first comprehensive landscape analysis of Web3-AI agent integration,
examining \nPopCount capitalized tokens worth \nPopMcap at two
timepoints \nWindow apart across landscape, finance, governance,
security and trust dimensions. AI agents acquire trustless execution
guarantees from Web3 infrastructure, while Web3 systems gain accessibility and
automation from AI.
Infrastructure dominates both project count (\nClsInfraCount of \nClsCount classified) and capitalization (\nClsInfraShare), though Financial Services grew disproportionately between snapshots. Valuation is extremely concentrated (Gini \nPopGini; top decile \nPopTopDecileShare) and declined \nCohChange across the archived cohort despite token population growth. On-chain standards show the same concentration pattern: x402 settles hundreds of thousands of machine payments with negligible per-transaction value, and ERC-8004 identity registries are populated but inflated by two structurally distinct patterns.
AI agents extend DeFi toward persistent automation, enable governance monitoring and analysis, strengthen security through automated vulnerability detection, and operate under cryptographic guarantees and decentralized verification. However, significant challenges persist in scalability, security, ethics, and technical integration. Our findings highlight four concrete research directions: agent memory persistence, portable digital assets, identity and attestation economics, and multi-agent coordination.